\documentclass[useAMS,usenatbib]{mn2e}
\usepackage{graphicx}

\def\Msun{M$_\odot$}

\def\Mgii{Mg\,{\sc ii}}
\def\Feii{Fe\,{\sc ii}}

\def\Nev{[Ne\,{\sc v}]}
\def\Neiii{[Ne\,{\sc iii}]}

\def\Oiii{[O\,{\sc iii}]}
\def\Oii{[O\,{\sc ii}]}
\def\Nii{[N\,{\sc ii}]}

\def\Ha{H$\alpha$}
\def\Hb{H$\beta$}

\def\kms{km s$^{-1}$}

\def\lsim{\mathrel{\rlap{\lower 3pt \hbox{$\sim$}} \raise 2.0pt \hbox{$<$}}}
\def\gsim{\mathrel{\rlap{\lower 3pt \hbox{$\sim$}} \raise 2.0pt \hbox{$>$}}}

\title[BH binary candidates: I. Spectral properties and evolution]{
The nature of massive black hole binary candidates: I. Spectral properties 
and evolution
}
\author[Decarli et al.]{Roberto Decarli$^{1}$\thanks{E-mail: decarli@mpia.de}, 
Massimo Dotti$^{2,3}$, Michele Fumagalli$^{4,5,6}$, Paraskevi Tsalmantza$^{7}$,   \newauthor 
Carmen Montuori$^{8,9}$, Elisabeta Lusso$^{1}$, David W. Hogg$^{1,10}$ and Jason X. Prochaska$^{11}$\\
$^{1}$Max-Planck Institut f\"{u}r Astronomie, K\"{o}nigstuhl 17, D-69117, Heidelberg, Germany\\
$^{2}$Dipartimento di Fisica G.~Occhialini, Universit\`a degli Studi di Milano Bicocca, Piazza della Scienza 3, 20126 Milano, Italy\\
$^{3}$INFN, Sezione di Milano-Bicocca, Piazza della Scienza 3, 20126 Milano, Italy\\
$^{4}$Carnegie Observatories, 813 Santa Barbara Street, Pasadena, CA 91101, USA. \\
$^{5}$Department of Astrophysics, Princeton University, Princeton, NJ 08544-1001, USA.\\
$^{6}$Hubble Fellow\\
$^{7}$Klausenpfad 22, Heidelberg, D-691221, Germany\\
$^{8}$Technion, Department of Physics, IL-32000, Haifa, Israel\\
$^{9}$Department of Physics, Faculty of Natural Sciences, University of Haifa, Haifa 31905, Israel\\
$^{10}$Center~for~Cosmology~and~Particle~Physics, Department~of~Physics, New~York~University, 4~Washington~Place, New~York, NY 10003, USA\\
$^{11}$Department of Astronomy and Astrophysics, University of California, 1156, High Street, Santa Cruz, CA 95064, USA\\
}

\begin{document}


\pagerange{\pageref{firstpage}--\pageref{lastpage}} \pubyear{2012}

\maketitle

\label{firstpage}

\begin{abstract}
Theoretically, bound binaries of massive black holes are expected as the 
natural outcome of mergers of massive galaxies. From the observational side, 
however, massive black hole binaries remain elusive. Velocity shifts 
between narrow and broad emission lines in quasar spectra
are considered a promising observational tool to search for spatially 
unresolved, dynamically bound binaries. In this 
series of papers we investigate the nature of such candidates through 
analyses of their spectra, images and multi-wavelength spectral 
energy distributions. Here we investigate the properties of the optical 
spectra, including the evolution of the broad line profiles, of all the 
sources identified in our previous study.
We find a diverse phenomenology of broad and narrow line luminosities, 
widths, shapes, ionization conditions and time variability,
which we can broadly ascribe to 4 classes based on the shape of the
broad line profiles: 1) Objects with bell-shaped broad lines with big 
velocity shifts ($>$1000 \kms) compared to their narrow lines show a 
variety of broad line widths and luminosities, modest flux variations
over a few years, and no significant change in the broad line peak
wavelength. 2) Objects with double-peaked broad emission lines tend
to show very luminous and broadened lines, and little time 
variability. 3) Objects with asymmetric broad emission lines show a broad
range of broad line luminosities and significant variability of the line 
profiles. 4) The remaining sources tend to show moderate to low broad line 
luminosities, and can be ascribed to diverse phenomena. We discuss the
implications of our findings in the context of massive black hole binary
searches.
\end{abstract}


\section{Introduction}

In the framework of hierarchical models of galaxy formation, mergers
play a key role in the build-up of galaxies. Massive black holes (BHs)
ubiquitously populate the centres of galaxies
\citep[at least in massive galaxies; e.g.][]{kormendy95,decarli07,gallo08}.
As a natural consequence, BH pairs wandering in merger remnants are 
expected. Dynamical friction efficiently drives the 
BHs towards the centre of the potential well, where they form a bound
black hole binary (BHB). The subsequent evolution of these systems is
poorly understood. At separations of a few parsecs, dynamical friction 
becomes inefficient, and the binary may stall \citep[the so-called 
`last parsec problem'; for recent reviews, see][]{colpi09,dotti12}. 
If this limit is overcome, and the binary reaches milli-parsec 
separations, then gravitational wave emission becomes
significant, and the binary rapidly shrinks towards the coalescence of 
the two BHs.

Examples of unbound BH pairs, with separations of 0.1--10 kpc, have
been observed, as in the prototypical case of NGC 6240
\citep{komossa03}, in Arp 299 \citep{ballo04}, in IRAS
20210+1121 \citep{piconcelli10}, in Mrk 463 \citep{bianchi08}, Mrk 739
\citep{koss11}, and NGC3393 \citep{fabbiano11}. Most of these systems
show strong evidence of on-going mergers involving gas-rich galaxies,
with dramatically perturbed host galaxy morphologies, intense star
formation rates and high FIR luminosities. However, little is known
about {\em bound} BHBs. The only spatially resolved BHB candidate to
date is 0402+379, a radio galaxy with two compact cores at a projected
separation of $\approx 7$ pc \citep{maness04,rodriguez06}. 

\begin{table*} 
\caption{{\rm Summary of the follow-up observations. (1) Quasar name. 
(2-3) Right ascension and declination (J2000).  (4) Object classification,
based on the shape of Balmer lines: B- BHB candidates; D- DPEs; A- sources with
Asymmetric line profiles; O- Others. (5) Redshift of narrow
lines. (6) Date of spectroscopic follow-up observations. (7) Time lag
between SDSS observations and our follow-up spectroscopy,
in the rest-frame of $z_{\rm NL}$. (8) Telescope: C2.2m=Calar Alto 2.2m; 
C3.5m=Calar Alto 3.5m; Lick; Keck. (9) Weather conditions during the 
spectroscopic follow-up: P=photometric; NP=not photometric; C=cloudy.
}} \label{tab_journal}
\begin{center}
\begin{tabular}{ccccccccc}
   \hline	      
    Obj.name    & R.A.    &  Dec.  & Class. & $z_{\rm NL}$& Date & $\Delta t_0$ & Tel. & Weather  \\
		&         &	   &        &             &      & [yr]         &      &          \\
     (1)	&   (2)   &  (3)   & (4)    &  (5)        & (6)  &  (7)         &  (8) &  (9)     \\
   \hline 
J0012-1022 & 00:12:24.03 & -10:22:26.3 & A  & 0.228 & 2011-07-22 & 8.08 & C3.5m & P  \\  
J0155-0857 & 01:55:30.02 & -08:57:04.0 & O  & 0.165 & 2011-09-26 & 8.62 & Lick  & C  \\  
J0221+0101 & 02:21:13.15 & +01:01:02.9 & O  & 0.354 & 2011-09-26 & 8.02 & Lick  & C  \\  
J0829+2728 & 08:29:30.60 & +27:28:22.7 & O  & 0.321 & 2012-02-22 & 6.32 & C2.2m & NP \\  
J0918+3156 & 09:18:33.82 & +31:56:21.2 & O  & 0.452 & 2012-02-20 & 5.64 & C2.2m & NP \\  
J0919+1108 & 09:19:30.32 & +11:08:54.0 & O  & 0.369 & 2012-02-20 & 5.86 & C2.2m & NP \\  
J0921+3835 & 09:21:16.13 & +38:35:37.6 & A  & 0.187 & 2012-02-22 & 7.50 & C2.2m & NP \\  
J0927+2943 & 09:27:12.65 & +29:43:44.1 & B  & 0.713 & 2012-02-20 & 4.14 & C2.2m & NP \\  
J0931+3204 & 09:31:39.05 & +32:04:00.2 & O  & 0.226 & 2012-02-21 & 5.79 & C2.2m & NP \\  
J0932+0318 & 09:32:01.60 & +03:18:58.7 & BD & 0.420 & 2012-02-22 & 7.19 & C2.2m & NP \\  
J0936+5331 & 09:36:53.85 & +53:31:26.9 & A  & 0.228 & 2011-03-15 & 7.48 & C2.2m & C  \\  
J0942+0900 & 09:42:15.12 & +09:00:15.8 & D  & 0.213 & 2012-02-21 & 7.28 & C2.2m & NP \\  
J0946+0139 & 09:46:03.95 & +01:39:23.7 & A  & 0.220 & 2012-02-22 & 8.96 & C2.2m & NP \\  
J1000+2233 & 10:00:21.80 & +22:33:18.6 & BD & 0.419 & 2011-01-03 & 3.53 & Keck  & NP \\  
J1010+3725 & 10:10:34.28 & +37:25:14.8 & O  & 0.282 & 2012-02-22 & 6.38 & C2.2m & NP \\  
J1012+2613 & 10:12:26.86 & +26:13:27.3 & BD & 0.378 & 2012-02-22 & 4.42 & C2.2m & NP \\  
J1027+6050 & 10:27:38.54 & +60:50:16.5 & D  & 0.332 & 2012-02-22 & 7.41 & C2.2m & NP \\  
J1050+3456 & 10:50:41.36 & +34:56:31.4 & B  & 0.272 & 2012-02-21 & 5.49 & C2.2m & NP \\  
J1105+0414 & 11:05:39.64 & +04:14:48.2 & D  & 0.436 & 2012-02-23 & 6.91 & C2.2m & NP \\  
J1117+6741 & 11:17:13.91 & +67:41:22.7 & O  & 0.248 & 2012-02-21 & 8.86 & C2.2m & NP \\  
J1154+0134 & 11:54:49.42 & +01:34:43.6 & BA & 0.469 & 2011-03-15 & 6.68 & C2.2m & C  \\  
J1207+0604 & 12:07:55.83 & +06:04:02.8 & O  & 0.136 & 2012-02-23 & 8.69 & C2.2m & NP \\  
J1211+4647 & 12:11:13.97 & +46:47:12.0 & O  & 0.294 & 2012-02-21 & 6.06 & C2.2m & NP \\  
J1215+4146 & 12:15:22.78 & +41:46:21.0 & O  & 0.196 & 2012-02-21 & 6.55 & C2.2m & NP \\  
J1216+4159 & 12:16:09.60 & +41:59:28.4 & O  & 0.242 & 2012-02-22 & 6.31 & C2.2m & NP \\  
J1328-0129 & 13:28:34.15 & -01:29:17.6 & O  & 0.151 & 2012-02-20 & 8.45 & C2.2m & NP \\  
J1414+1658 & 14:14:42.03 & +16:58:07.2 & O  & 0.237 & 2012-02-23 & 3.23 & C2.2m & NP \\  
J1440+3319 & 14:40:05.31 & +33:19:44.5 & A  & 0.179 & 2012-02-23 & 5.77 & C2.2m & NP \\  
J1536+0441 & 15:36:36.22 & +04:41:27.0 & BD & 0.389 & 2012-02-23 & 2.78 & C2.2m & NP \\  
J1539+3333 & 15:39:08.09 & +33:33:28.0 & B  & 0.226 & 2011-03-12 & 6.28 & C2.2m & C  \\  
J1652+3123 & 16:52:55.90 & +31:23:43.8 & O  & 0.593 & 2011-07-23 & 5.12 & C3.5m & P  \\  
J1714+3327 & 17:14:48.51 & +33:27:38.3 & BO & 0.181 & 2011-07-21 & 2.72 & C3.5m & P  \\  
\hline
 \end{tabular}
 \end{center}
\end{table*}

Recently, a number of BHB candidates have been identified in the 
Sloan Digital Sky Survey \citep[SDSS;][]{york00} spectroscopic 
database based on large velocity shifts ($>1000$ \kms{}) between 
narrow and broad emission lines (NLs and BLs, respectively). In the 
BHB scenario, under the assumption that one of the two BHs is active,
the BLs are shifted with respect to the host galaxy
rest-frame (as traced by the NLs) as a consequence of the Keplerian
motion of the binary \citep{begelman80}. A handful of BHB candidates 
have been found in this way: SDSS J092712.65+294344
\citep{komossa08,bogdanovic09,dotti09}, SDSS J153636.22+044127
\citep{boroson09}, SDSS J105041.35+345631.3 \citep{shields09a}, 
4C+22.25 \citep[J1000+2233 in this paper;][]{decarli_4c2225}, and 
SDSS J093201.60+031858.7 \citep{barrows11}. More recently, we 
performed the first systematic search for objects with these properties 
in the SDSS. We used a data-driven method -- heteroscedastic matrix 
factorization, HMF -- developed within our group \citep{vivi12}. 
The method has been successfully applied in order to identify objects
with spectroscopic features associated with two different redshift 
systems. Specifically, we used HMF in order to model spectra of 
$\sim55000$ quasars and $\sim4000$ galaxies and to highlight objects 
with shifted BLs with respect to the NLs. This search produced a list
of 32 interesting objects \citep{vivi11}. A similar, systematic 
search has been independently performed by \citet{eracleous12}.
In this case, interesting objects were selected on the basis of 
anomalous profiles of the broad \Hb{} line, identified as outliers
of Principal Component Analysis fits of the spectra. This yielded 
88 candidates, 13 of which in common with the sample defined in 
\citet{vivi11}. This relatively small overlap is likely due to the
different approaches followed in the two studies: the method used in
\citet{vivi11} aims at discovering objects with multiple redshift 
components, while the work by \citet{eracleous12} identifies sources
with peculiar \Hb{} line profiles 
\citep[for a more detailed comparison between these two studies, see Sec.\,3 of][]{eracleous12}.

The spectroscopically-selected BHB candidates discussed in \citet{vivi11}
show a variety of broad 
line shapes, widths, luminosities, etc. This suggests that the 
nature of these objects is intrinsically heterogeneous. Besides the BHB
scenario, velocity shifts
between broad and narrow emission lines can be attributed to peculiar 
properties of the BL region of a single AGN, such as non-axisymmetric
perturbations in the BL region emissivity (similar to `hot spots' in 
the accretion disc), or to
an edge-on disc-like geometry of the region emitting BLs, as observed in 
Double-Peaked Emitters \citep[hereafter, DPEs; see e.g.][]{strateva03}. 
Alternatively, BLs and NLs may arise from two distinct sources, aligned 
along the line of sight. Finally, shifted BLs can be attributed to a 
recoiling massive BH, resulting from the coalescence of a BHB 
\citep[see, e.g.,][]{komossa08}. 

\begin{figure}
\includegraphics[width=0.99\columnwidth]{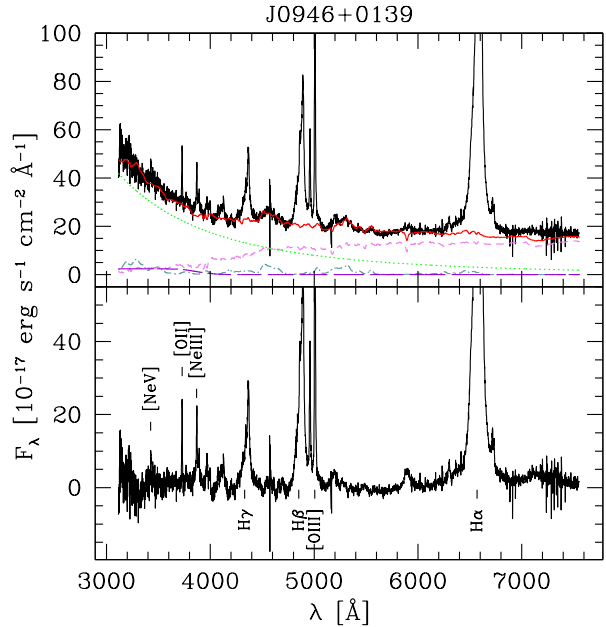}
\caption{Example of continuum decomposition. {\em Top panel:} SDSS
spectrum of J0946+0139. Dotted, dash-dotted, short-dashed, and
long-dashed lines show the power-law, the \Feii{} template, the host galaxy
template, and the Balmer pseudo-continuum. The total fit is also shown
as a red, solid line. {\em Bottom panel:} Residual spectrum. Main emission
lines are also labeled.
}
\label{fig_contfit}
\end{figure}
\begin{figure}
\includegraphics[width=0.99\columnwidth]{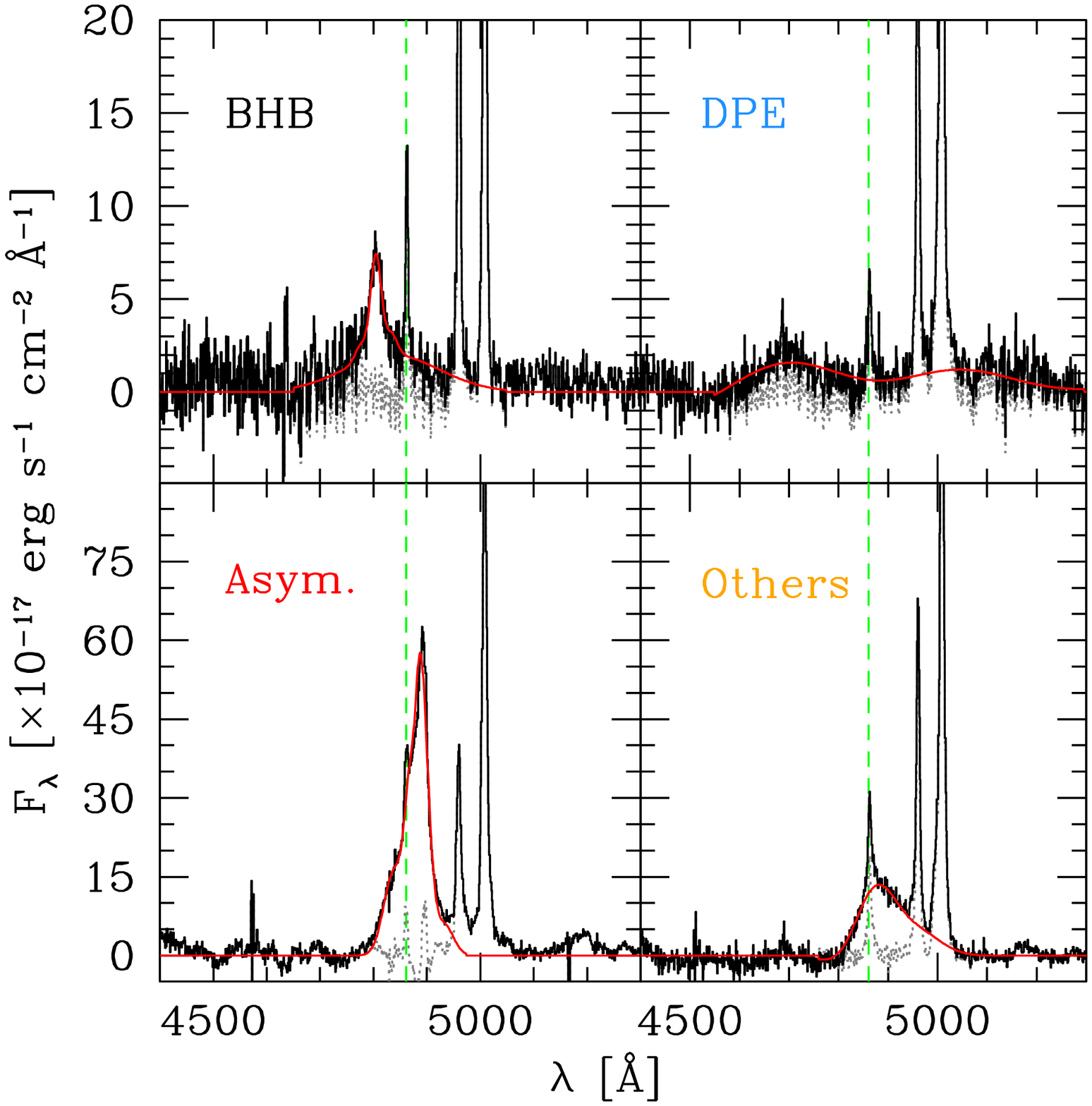}
\caption{Examples of the broad line profiles in each class of objects in our
sample. We label as `BHB candidates', objects with fairly symmetric broad 
line profiles, with clear shifts ($>$1000 \kms{}) with respect to their 
narrow components (here: J1050+3456); as `DPE candidates', objects which 
show a double-horned profile of broad lines (in this figure: J1000+2233); as
`Asymmetric', objects in which the broad lines have a symmetric base but a 
highly skewed, asymmetric core (here: J0946+0139); and as `Others', sources
with small shifts or irregular line profiles (here: J1414+1658). In the 
plots, we show the observed SDSS spectra around the \Hb{} region (solid 
black line), after the continuum emission has been modeled out and 
subtracted (see text for details) and after shifting the spectra back to the
rest frame (based on $z_{\rm NL}$; see Table \ref{tab_journal}). The dashed 
vertical lines mark the narrow component of \Hb{}. The red lines show the 
fit to the broad component of \Hb{}. Residuals are plotted with grey, dotted
lines.
}
\label{fig_profiles}
\end{figure}

The purpose of this series of papers is to test these scenarios against
the BHB hypothesis for all the 32 sources in the sample identified by
\citet{vivi11}. In particular, here we discuss their spectral properties
(fluxes, luminosities, widths, broad line profiles and evolution). In  
paper II (Lusso et al., in prep.), we study their multi-wavelength Spectral 
Energy Distributions. Finally, in paper III (Decarli et al., in prep.) we
analyse the images of their host galaxies and their galactic environment. 

The structure of this paper is the following: in Section 
\ref{sec_data} we describe the follow-up observations and the data 
reduction. The analysis of the SDSS spectra and of our second-epoch data
are presented in Sections \ref{sec_spectra} and \ref{sec_evol} respectively. 
Discussion and conclusions are drawn in Section \ref{sec_conclusions}.
Throughout the paper we will assume a standard cosmology with 
$H_0=70$ km s$^{-1}$ Mpc$^{-1}$, $\Omega_{\rm m}=0.3$ and 
$\Omega_{\Lambda}=0.7$.

\section{Observational data set}\label{sec_data}

Our data set consists of SDSS spectra of the 32 sources
identified in \citet{vivi11} and of dedicated follow-up long-slit optical
spectra. SDSS spectra cover the 3800--9200 \AA{} wavelength
range with a spectral resolution of $\lambda/\Delta\lambda\approx1800$
\citep{york00,sdssdr7}. SDSS fibers have an aperture diameter of $3''$ on
sky. 

Follow-up observations were carried out at various facilities: Calar Alto 
2.2m and 3.5m, Lick Observatory, and Keck. The journal of observations is 
presented in Table \ref{tab_journal}.

\begin{table*}
\caption{{\rm Line luminosities of \Nev{}, \Neiii{}, \Oii{}, \Oiii{}, broad \Hb{} and \Ha{} as measured from the SDSS spectra.
}} \label{tab_lines}
\begin{center}
\begin{tabular}{ccccccc}
   \hline	      
    Obj.name    & $L$ (\Nev) & $L$ (\Neiii) & $L$ (\Oii) & $L$ (\Oiii) &  $L$ (\Hb{}, B) & $L$ (\Ha{}, B)  \\ 
		& [$10^{41}$ erg\,s$^{-1}$]  & [$10^{41}$ erg\,s$^{-1}$]    & [$10^{41}$ erg\,s$^{-1}$] & [$10^{41}$ erg\,s$^{-1}$] & [$10^{41}$ erg\,s$^{-1}$] & [$10^{41}$ erg\,s$^{-1}$] \\ 
     (1)	&   (2)   &  (3)   & (4)        &  (5) & (6)          &  (7) \\ 
   \hline 
  J0012-1022 & $ 0.34\pm0.35 $ & $ 1.46\pm0.19 $ & $ 1.97\pm0.21 $ & $  7.79\pm0.19 $ & $ 34.59\pm 3.46$ & $ 96.84\pm3.46 $ \\	
  J0155-0857 & $ 1.12\pm0.21 $ & $ 2.03\pm0.14 $ & $ 1.42\pm0.15 $ & $  4.83\pm0.19 $ & $ 32.25\pm 3.22$ & $ 92.28\pm3.22 $ \\	
  J0221+0101 & $ 2.59\pm0.32 $ & $ 2.12\pm0.21 $ & $ 1.73\pm0.22 $ & $  6.70\pm0.22 $ & $ 25.18\pm 2.52$ & $ 54.18\pm2.52 $ \\	
  J0829+2728 & $ 3.03\pm0.33 $ & $ 3.77\pm0.23 $ & $ 3.64\pm0.25 $ & $  9.97\pm0.30 $ & $ 16.54\pm 1.65$ & $ 49.60\pm1.65 $ \\	
  J0918+3156 & $ 2.82\pm0.56 $ & $10.22\pm0.58 $ & $16.32\pm0.67 $ & $ 38.80\pm0.84 $ & $124.46\pm12.45$ &       ---        \\	
  J0919+1108 & $15.24\pm1.71 $ & $ 8.48\pm1.11 $ & $ 2.41\pm0.53 $ & $ 22.75\pm0.59 $ & $103.43\pm10.34$ &       ---        \\	
  J0921+3835 & $ 1.31\pm0.39 $ & $ 0.27\pm0.11 $ & $ 1.10\pm0.14 $ & $  7.57\pm0.17 $ & $ 17.93\pm 1.79$ & $ 51.76\pm1.79 $ \\	
  J0927+2943 & $ 2.91\pm2.26 $ & $ 7.51\pm1.04 $ & $23.25\pm1.23 $ & $ 35.44\pm2.20 $ & $ 93.99\pm 9.40$ &       ---        \\	
  J0931+3204 & $ 2.77\pm0.28 $ & $ 1.39\pm0.11 $ & $ 0.33\pm0.08 $ & $  5.62\pm0.14 $ & $  4.94\pm 0.49$ & $ 28.26\pm0.49 $ \\	
  J0932+0318 & $ 4.60\pm0.77 $ & $ 3.95\pm0.56 $ & $ 4.66\pm0.49 $ & $ 13.30\pm0.49 $ & $ 39.95\pm 3.99$ &       ---        \\	
  J0936+5331 & $ 2.02\pm0.32 $ & $ 3.10\pm0.20 $ & $ 0.55\pm0.15 $ & $  7.76\pm0.22 $ & $ 58.87\pm 5.89$ & $163.95\pm5.89 $ \\	
  J0942+0900 & $ 2.29\pm0.37 $ & $ 3.69\pm0.21 $ & $ 5.43\pm0.27 $ & $ 14.93\pm0.30 $ & $ 24.46\pm 2.45$ & $ 62.86\pm2.45 $ \\	
  J0946+0139 & $ 0.98\pm0.12 $ & $ 3.45\pm0.11 $ & $ 2.03\pm0.10 $ & $  7.48\pm0.11 $ & $ 30.55\pm 3.05$ & $101.07\pm3.05 $ \\	
  J1000+2233 & $ 7.86\pm0.65 $ & $ 3.74\pm0.37 $ & $ 7.24\pm0.42 $ & $ 20.58\pm0.58 $ & $ 23.91\pm 2.39$ &       ---        \\	
  J1010+3725 & $ 0.50\pm0.35 $ & $ 4.31\pm0.22 $ & $ 6.35\pm0.26 $ & $ 34.76\pm0.32 $ & $ 16.47\pm 1.65$ &       ---        \\	
  J1012+2613 & $ 5.47\pm0.48 $ & $ 4.56\pm0.39 $ & $ 3.86\pm0.26 $ & $ 21.08\pm0.37 $ & $ 20.40\pm 2.04$ &       ---        \\	
  J1027+6050 & $ 4.48\pm1.05 $ & $ 0.80\pm0.42 $ & $ 2.45\pm0.49 $ & $  7.10\pm0.44 $ & $ 49.53\pm 4.95$ & $ 80.24\pm4.95 $ \\	
  J1050+3456 & $ 0.72\pm0.25 $ & $ 1.32\pm0.16 $ & $ 3.32\pm0.19 $ & $  4.33\pm0.24 $ & $  7.57\pm 0.76$ & $ 26.63\pm0.76 $ \\	
  J1105+0414 & $ 4.65\pm0.72 $ & $ 3.66\pm0.52 $ & $ 4.06\pm0.63 $ & $ 22.84\pm0.66 $ & $109.16\pm10.92$ &       ---        \\	
  J1117+6741 & $ 2.35\pm0.39 $ & $ 0.88\pm0.17 $ & $ 2.54\pm0.13 $ & $  4.92\pm0.12 $ & $  3.36\pm 0.34$ & $ 16.05\pm0.34 $ \\	
  J1154+0134 & $ 0.65\pm0.96 $ & $ 2.08\pm0.73 $ & $ 9.73\pm0.84 $ & $ 15.47\pm0.78 $ & $ 52.47\pm 5.25$ &       ---        \\	
  J1207+0604 & $ 0.74\pm0.09 $ & $ 0.41\pm0.04 $ & $ 0.30\pm0.05 $ & $  1.72\pm0.06 $ & $  4.20\pm 0.42$ & $ 12.78\pm0.42 $ \\	
  J1211+4647 & $ 3.56\pm0.44 $ & $ 3.47\pm0.28 $ & $ 1.35\pm0.21 $ & $  9.62\pm0.23 $ & $ 19.71\pm 1.97$ & $ 78.66\pm1.97 $ \\	
  J1215+4146 & $ 0.54\pm0.16 $ & $ 1.83\pm0.10 $ & $ 3.26\pm0.12 $ & $ 11.87\pm0.18 $ & $  1.36\pm 0.14$ & $ 32.68\pm0.14 $ \\	
  J1216+4159 & $ 1.26\pm0.16 $ & $ 2.44\pm0.16 $ & $ 3.18\pm0.19 $ & $  6.96\pm0.23 $ & $  4.58\pm 0.46$ & $ 28.09\pm0.46 $ \\	
  J1328-0129 & $ 0.93\pm0.09 $ & $ 1.14\pm0.06 $ & $ 1.37\pm0.06 $ & $  3.64\pm0.11 $ & $  3.79\pm 0.38$ & $ 22.33\pm0.38 $ \\	
  J1414+1658 & $ 3.61\pm0.25 $ & $ 5.91\pm0.17 $ & $ 3.15\pm0.15 $ & $ 15.37\pm0.19 $ & $ 17.94\pm 1.79$ & $ 65.20\pm1.79 $ \\	
  J1440+3319 & $ 0.47\pm0.16 $ & $ 0.46\pm0.08 $ & $ 1.21\pm0.10 $ & $  3.33\pm0.11 $ & $  3.15\pm 0.32$ & $ 16.44\pm0.32 $ \\	
  J1536+0441 & $ 4.11\pm0.64 $ & $ 5.89\pm0.51 $ & $ 3.05\pm0.61 $ & $ 26.87\pm0.73 $ & $ 95.49\pm 9.55$ &       ---        \\	
  J1539+3333 & $ 0.12\pm0.16 $ & $ 0.01\pm0.10 $ & $ 1.13\pm0.12 $ & $  0.37\pm0.09 $ & $  1.56\pm 0.16$ & $  4.56\pm0.16 $ \\	
  J1652+3123 & $ 9.55\pm1.85 $ & $17.38\pm1.42 $ & $14.41\pm1.51 $ & $ 59.45\pm1.90 $ & $208.83\pm20.88$ &       ---        \\	
  J1714+3327 & $ 0.81\pm0.22 $ & $ 3.53\pm0.14 $ & $ 0.80\pm0.13 $ & $ 11.59\pm0.15 $ & $ 27.48\pm 2.75$ & $ 88.43\pm2.75 $ \\	
  \hline
 \end{tabular} 
 \end{center}
\end{table*}

The bulk of the spectroscopic follow-up has been performed using the Calar
Alto Faint Object Spectrograph (CAFOS) on the 2.2m telescope in Calar Alto.
Observations were carried out with the g100 or r100 grisms, which allow a 
spectral resolution 
$\lambda/\Delta\lambda\approx700$ (with $1''$ slit) as measured from the 
profile of sky emission lines. The observed ranges in the two setups are 
4900--7800 \AA{} and 5900--9000 \AA{} for the g100 and r100 grisms 
respectively.
Four sources (J0012-1022, J0927+2943, J1652+3123 and J1714+3327) were 
observed with the Cassegrain Twin Spectrograph (TWIN) at the 3.5m telescope 
in Calar Alto. The T07 grism has been adopted on the red arm, providing 
continuous spectral coverage in the range 5500--11000 \AA{} with a spectral 
resolution $\lambda/\Delta\lambda\approx1300$ ($1''$ slit).
Two targets (J0155-0857 and J0221+0101) have been observed at the Lick 
observatory using the Kast spectrograph. Due to instrumental problems, 
data were collected only on the red side using the 600/7500 grating, the 
D55 beam splitter, and the $2"$ slit.  The two objects were observed for 
1800s in cloudy conditions and at airmasses of 1.5 and 1.3, 
respectively. A spectrum of the standard star G191B2B was also obtained 
with the same instrument configuration for flux calibration. Finally, 
one object, J1000+2233, has been observed while transiting at airmass of 
1.3 in non-photometric conditions with the LRIS/Keck spectrograph \citep{oke95}. The 
spectrograph was configured with the 400/3400 grism, the 1200/7500 
grating blazed at 9200 \AA{}, $1''$ slit, and the D560 dichroic. Using the 
2$\times$2 binning, this configuration yields a final spectral resolution of 
2.1 \AA{}/pix in the blue and 0.8 \AA{}/pix in the red. Two exposures of 600s 
each were acquired. The spectro-photometric star HZ44 was also observed 
with the same setup for flux calibration.

Standard \textsf{IRAF} tools were used in the reduction of Calar Alto data, 
in order to create master bias and flat field images and to correct for them, 
to align and combine frames, perform wavelength and relative flux calibration, 
to perform background subtraction, and to extract 1D spectra. Multiple
(3-6) frames were collected in order to easily remove cosmic rays. Sky
emission lines were used in order to perform wavelength
calibration. This allows us to take into account instrument
flexures. The accuracy of the wavelength calibration is typically of
$\lsim1$ \AA{} (residual rms) over the whole observed range. 
This corresponds to minimum velocity uncertainties of $\sim40$ \kms{}, if we 
observe a line at 5000 \AA{} (rest frame) shifted to a fiducial redshift 
$z=0.4$ (much smaller than typical uncertainties in the fit of broad line
peak wavelengths, see Section \ref{sec_evol}). Relative flux calibration 
was performed using spectra of spectrophotometric standards, e.g., Feige56, 
Feige66, BD+75$^\circ$\,325, BD+33$^\circ$\,2642. A similar
procedure was adopted during the reduction of the Lick and Keck data
using the LowRedux pipeline
(http://www.ucolick.org/$\sim$xavier/LowRedux/index.html). Since
observations were mostly performed under non-photometric conditions,
absolute flux calibration is uncertain. Since we are mostly interested
in the evolution of the profiles of broad emission lines, we scaled
the spectra so that the fluxes of the \Oiii{} and \Nii{} narrow lines
in our spectra match the ones observed in the SDSS data. The median
correction corresponds to a factor $\sim 1.4$.

\section{Spectral properties}\label{sec_spectra}

In order to study the emission line properties of these sources, we model
and subtract the continuum emission following a procedure similar to the one
presented in \citet{decarli10a} and \citet{derosa11}. Namely, the continuum 
emission is fitted with a superposition of a power-law, a Balmer 
pseudocontinuum \citep{grandi82}, a host galaxy model \citep[here we refer
to the templates by][]{kinney96}, and a \Feii{} template \citep[adapted
from][]{verner09}. An example of continuum modeling is shown in Figure 
\ref{fig_contfit}.

\begin{figure}
\includegraphics[width=0.99\columnwidth]{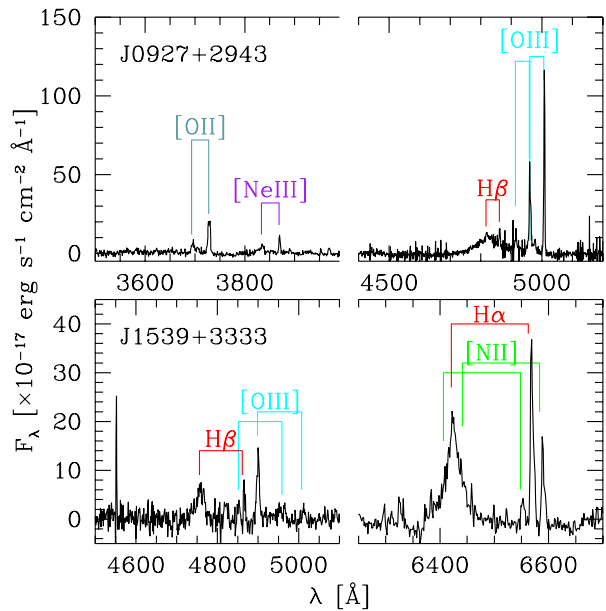}
\caption{Continuum-subtracted SDSS spectra of J0927+2943 and J1539+3333.
Both spectra show two sets of narrow lines, the most prominent of which
are labeled.}
\label{fig_nl}
\end{figure}

\begin{table}
\caption{{\rm Line widths of \Oii{}, \Oiii{} and of the broad component of \Hb{} in the targets of our sample.
}} \label{tab_fwhm}
\begin{center}
\begin{tabular}{cccc}
   \hline	      
    Obj.name    & \multicolumn{3}{c}{FWHM [\kms]} \\
		& \Oii & \Oiii & \Hb{}, B \\
     (1)	&   (2)  &  (3)   & (4)      \\
   \hline 
%
  J0012-1022 &  400 &  480 &  4500  \\  		 
  J0155-0857 &  440 &  300 &  7300  \\  		 
  J0221+0101 &  440 &  330 &  7100  \\  		 
  J0829+2728 &  440 &  420 &  5300  \\  		 
  J0918+3156 &  640 &  570 &  6700  \\  		 
  J0919+1108 &  560 &  840 &  7900  \\  		 
  J0921+3835 &  480 &  600 &  3100  \\  		 
  J0927+2943 &  440 &  240 &  6200  \\  		 
  J0931+3204 &  320 & 1050 &  3800  \\  		 
  J0932+0318 &  480 &  480 & 11800  \\  		 
  J0936+5331 &  400 &  540 &  4700  \\  		 
  J0942+0900 &  480 &  570 & 22000  \\  		 
  J0946+0139 &  520 &  570 &  2700  \\  		 
  J1000+2233 &  480 &  420 & 23000  \\  		 
  J1010+3725 &  640 & 1140 &  2000  \\  		 
  J1012+2613 &  480 &  630 & 19000  \\  		 
  J1027+6050 &  400 &  360 & 29000  \\  		 
  J1050+3456 &  480 &  300 &  2900  \\  		 
  J1105+0414 &  560 &  510 & 27000  \\  		 
  J1117+6741 &  560 &  480 &  7100  \\  		 
  J1154+0134 &  480 &  570 &  4200  \\  		 
  J1207+0604 &  440 &  330 & 11400  \\  		 
  J1211+4647 &  480 &  600 &  5400  \\  		 
  J1215+4146 &  640 &  750 &  3600  \\  		 
  J1216+4159 &  480 &  420 & 13000  \\  		 
  J1328-0129 &  560 &  360 & 10600  \\  		 
  J1414+1658 &  480 &  480 &  6900  \\  		 
  J1440+3319 &  560 &  390 &  6900  \\  		 
  J1536+0441 &  640 &  600 &  2300  \\  		 
  J1539+3333 &  480 &  420 &  1300  \\  		 
  J1652+3123 &  400 &  660 & 10100  \\  		 
  J1714+3327 &  560 &  660 &  4800  \\  		 
\hline
 \end{tabular}
 \end{center}
\end{table}

\begin{figure}
\includegraphics[width=0.99\columnwidth]{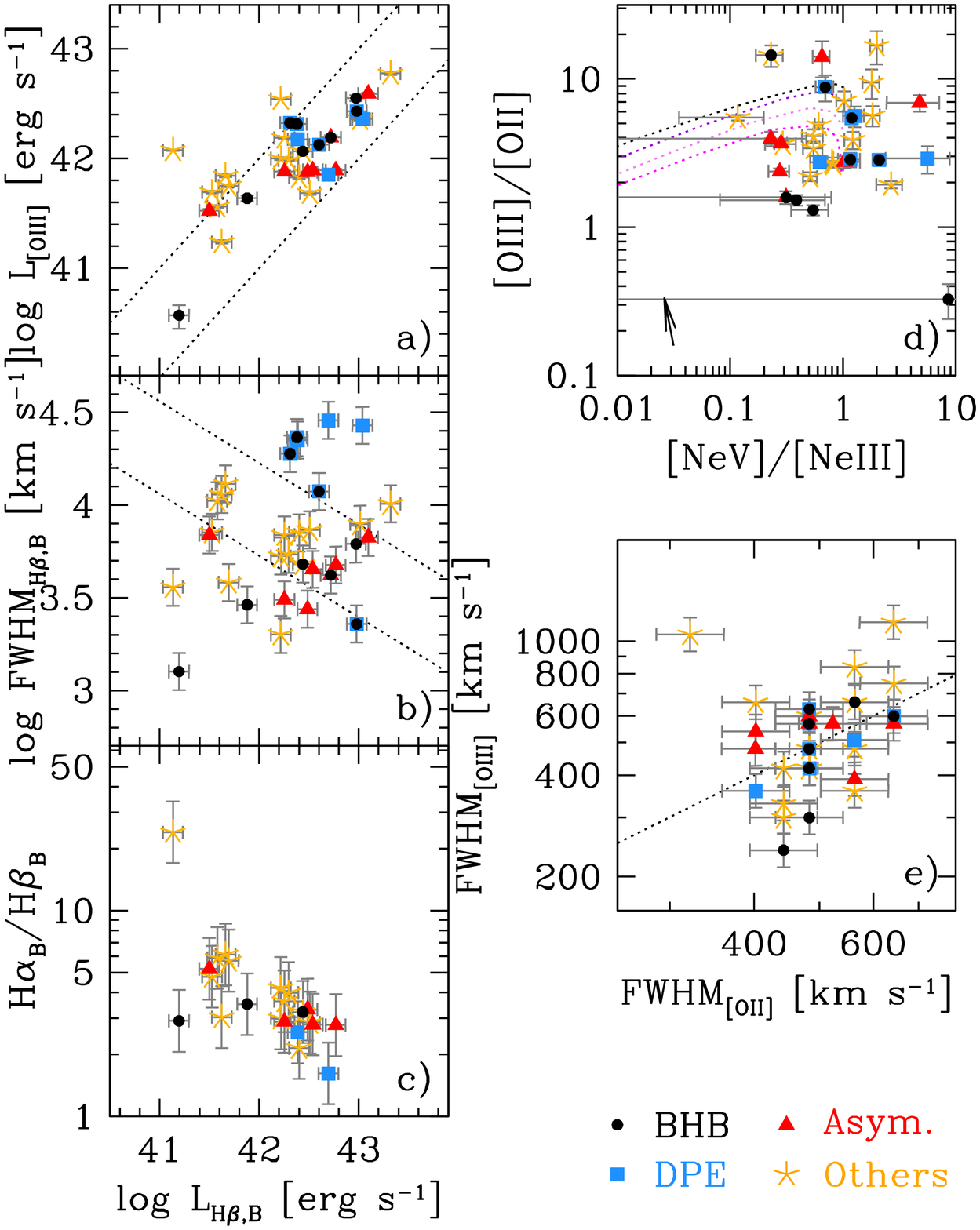}
\caption{Narrow and broad line diagnostics. Black, blue, red and yellow 
symbols mark BHB candidates, DPEs, objects with asymmetric line profiles 
and other sources, respectively. {\bf a)} Comparison between \Hb{} (broad)
and \Oiii{} luminosities. The dotted lines show the cases $L$(\Hb)=1--10 
$L$(\Oiii). {\bf b)} Width vs luminosity of broad \Hb{}. Dotted lines 
mark the loci of virial black hole masses equal to $10^9$ and $10^8$ 
M$_\odot$. {\bf c)} Flux ratio between the broad components of \Ha{} 
and \Hb{}, as a function of $L$(\Hb). 
{\bf d)} \Oiii/\Oii{} vs \Nev{}/\Neiii{} diagnostic of the gas ionization
in the interstellar medium. Dotted lines show the expected values 
for AGN-like ionization conditions at various spectral indexes 
(different curves; from dark to bright colours at increasing
spectral steepness) and ionization parameters (increasing from left 
to right). The arrow shows the effects of $A_V=2$ mag reddening. 
Adapted from \citet{groves04a,groves04b}. {\bf e)} Comparison
between the widths of \Oiii{} and \Oii{} (considered as a single line, as
the doublet cannot be resolved in any of these sources). The dotted line 
shows the 1-to-1 case. 
}
\label{fig_linediag}
\end{figure}

A variety of line profiles were reported in our original SDSS-based
selection. We introduce the following basic classification
scheme: fairly bell-shaped, strongly shifted broad lines (e.g., in J0927+2943, 
J1539+3333, J1714+3327) identify good BHB candidates \citep[although the 
BHB scenario does not necessarily imply a gaussian line profile, see 
e.g.][]{bogdanovic08,shen10}; very broad lines with tentative evidence of 
double-horned profiles are labeled as Double-Peaked Emitters 
(DPEs; e.g., in J1012-2613, J1027+6050, J1105+0414); lines showing a rather
symmetric base, centred at the redshift of the narrow lines, but an 
asymmetric core, resulting in a shifted peak, are called `Asymmetric' 
(e.g., in J0012-1022, J0936+5331, J0946+0139); and other, more complex 
profiles (e.g., J1010+3725, J1215+4146, etc) or lines with relatively small 
velocity shifts, or asymmetric wings but modest peak shift are referred to 
as `Others'.
Figure \ref{fig_profiles} shows a prototypical example for each class of 
objects\footnote{We note that this classification scheme is not univocal.
The ambiguity may be observational (poor S/N, or line wings overlapping onto
other spectral features) or due to the fact that the classes used here are 
not mutually exclusive, e.g., some DPEs have broad line profiles similar to 
the `Asymmetric' ones. }.


We focus on the properties of \Hb{}, \Oii{}, \Oiii{}, \Neiii{} and \Nev{}, 
which are all covered in the SDSS spectra of all the sources in our sample,
and of \Ha{} when available. 
The \Hb{} broad component is modeled with a superposition of a series of 
Gauss-Hermite polynomials (up to the seventh order), plus a gaussian curve. 
These components are required in order to 
reproduce the complex line profiles of the sources in our sample.
Examples of line fits are shown in Figure \ref{fig_profiles}. For all the 
other lines, we compute fluxes and peak wavelengths as zeroth and
first moments of the observed lines, and the line widths as Full Widths
at Half Maximum (FWHM) directly measured on the observed line profiles. 
Typical uncertainties are $\sim0.1$ dex and $\sim0.15$ dex in the luminosity
and widths of broad lines (mostly due to model degeneracies and fit residuals),
and of $\sim0.05$ dex in the widths of narrow lines (dominated by the noise
in the data). Tables \ref{tab_lines} and \ref{tab_fwhm} summarize the 
properties of these lines in each source.

The sources in our sample have been selected because they show two redshift
systems at $z_{\rm NL}$ and $z_{\rm BL}$. In 20 cases out of 32, $z_{\rm NL}$
$<$ $z_{\rm BL}$\footnote{The small asymmetry between the number of blue and 
redshifted BLs may be explained by the contamination of the \Oiii{} lines, 
which may hinder the identification of redshifted, weak broad \Hb{} emission 
lines}. Narrow \Oiii{} emission is 
observed at both $z_{\rm NL}$ and $z_{\rm BL}$ in only two cases, J0927+2943
and J1539+3333, both of which are considered BHB candidates in this analysis
(see Figure \ref{fig_nl}). 
The former source shows also \Oii{}, \Neiii{} and \Nev{} emission at both
redshifts, implying AGN-like ionization conditions of the gas emitting
the two sets of lines \citep[see also][]{komossa08}. For J1539+3333, no 
clear indication of \Nev{} is observed at any redshift \citep[for a 
discussion on how a second system of narrow lines can be interpreted in the 
light of the BHB hypothesis, see][]{bogdanovic09,dotti09}. 
Finally, the \Oiii{} lines of J1010+3725
show broad profiles with multiple components at various velocities, likely 
caused by massive gaseous outflows driven either by star formation or AGN 
winds. Such a feature is not observed in other, lower-ionization lines 
(e.g., \Oii{}). All the remaining sources show only one set of NLs. 
High-ionization lines (including \Neiii{} and \Nev{}) are ubiquitously reported.

Figure \ref{fig_linediag} shows various gas diagnostics based on
the luminosities and widths of both NLs and BLs. The luminosities
of \Oiii{} and \Hb{} (broad) are highly correlated, as generally observed
in normal quasars \citep{shen11}, with $L$(\Hb{},B)=1--10 $\times$ 
$L$(\Oiii{}). Largest deviations (e.g., the clear outlier J1215+4145) seem 
to be associated with high \Ha{},B/\Hb{},B ratios. DPEs in our
sample tend to show bright \Oiii{} and broad \Hb{} emission ($L$(\Hb,B) 
$>10^{42}$ erg\,s$^{-1}$),
with line widths $>$8000 \kms{}. Given that $L$(\Hb,B) is proportional to
the continuum luminosity, which is a proxy of the size of the BL region
\citep{bentz10}, this implies that, at face value, virial estimates of the 
BH mass in DPEs are large \citep[$>10^9$ M$_\odot$; for a discussion on the 
limitations of these estimates, see][]{wu04,lewis06}. 
Asymmetric broad lines tend to show bright BLs with $L$(\Hb,B) $>10^{42}$
erg\,s$^{-1}$ and intermediate to high $L$(\Hb,B)/$L$(\Oiii) ratios, 
and relatively modest (2500--5000 \kms) widths. The only exception is 
J1440+3319, which shows significantly lower BL luminosity 
($L$(\Hb,B)=$3.2\times10^{41}$ erg\,s$^{-1}$) but high \Ha,B/\Hb,B ($5.2$ instead of
$\sim3$ as observed in all the other asymmetric objects). BHB candidates 
and sources listed as `Others'
show a broad range of broad line luminosities and widths.
The majority (27/32) of our sources show typical AGN-like ionization
conditions \citep[see, f.i.,][]{groves04a,groves04b}, as traced by the 
\Oiii{}/\Oii{} vs \Nev{}/\Neiii{} diagnostic 
(due to its high ionization energy, \Nev{} cannot be 
photoionized by star formation only). A handful of sources
show \Nev{}/\Neiii{} values exceeding unity, which is not typically 
observed in lower luminosity AGN \citep{groves04a,groves04b}. This 
suggests a complete ionization of the clouds in the NL region. Among
the targets without clear AGN-like ionization conditions, 
three are BHB candidates (J1050+3456, J1154+0134 and J1539+3333), one is 
an object with asymmetric line profiles (J0012-1022) and one was 
classified as `Others' (J1010+3725). Finally, when comparing the FWHM
values of \Oiii{} and \Oii{}, we find a wide range of NL widths, from
practically unresolved (the BHB candidates J0927+2943 and J1050+3456,
or the source J0155-0857) to extremely broad (e.g., J0931+3204 and
J1010+3725, with FWHM$_{\rm [OIII]}$ $>$ 1000 \kms{}).

\section{Broad line temporal evolution}\label{sec_evol}

\begin{table}
\caption{{\rm Evolution of the shift of broad line peaks between 
SDSS and our follow-up observations, and corresponding (rest-frame) 
mean acceleration.
}} \label{tab_evo}
\begin{center}
\begin{tabular}{ccc}
   \hline	      
    Obj.name    & $\Delta v_{\rm BL}$ & Acceleration \\
		& [\kms{}] & [\kms{} yr$^{-1}$] \\
     (1)	&   (2)    &  (3)         \\
   \hline 
%
  J0012-1022 & $  -260_{ -210}^{ +130}$ & $  -32_{ -26}^{ +17}$  \\  
  J0155-0857 & $ -2500_{ -600}^{+3100}$ & $ -295_{ -70}^{+360}$  \\  
  J0221+0101 & $ +2300_{-5300}^{ +300}$ & $ +290_{-660}^{ +30}$  \\  
  J0829+2728 & $  -720_{ -520}^{ +620}$ & $ -110_{ -80}^{+100}$  \\  
  J0918+3156 & $ -1900_{ -800}^{+5200}$ & $ -340_{-140}^{+920}$  \\  
  J0919+1108 & $  -900_{ -500}^{ +500}$ & $ -160_{ -90}^{ +80}$  \\  
  J0921+3835 & $   +50_{ -300}^{ +420}$ & $  +10_{ -40}^{ +60}$  \\  
  J0927+2943 & $  -100_{ -500}^{ +500}$ & $  -20_{-120}^{+110}$  \\  
  J0931+3204 & $  -240_{ -230}^{ +220}$ & $  -40_{ -40}^{ +40}$  \\  
  J0932+0318 & $ +2600_{-2100}^{+1400}$ & $ +360_{-290}^{+200}$  \\  
  J0936+5331 & $ -2600_{ -100}^{+1600}$ & $ -350_{ -10}^{+220}$  \\  
  J0942+0900 & $  +400_{-3800}^{+1100}$ & $  +50_{-520}^{+140}$  \\  
  J0946+0139 & $  -260_{ -210}^{ +220}$ & $  -30_{ -25}^{ +25}$  \\  
  J1000+2233 & $ +3600_{-2500}^{+1100}$ & $+1000_{-700}^{+300}$  \\  
  J1010+3725 & $     0_{ -350}^{ +130}$ & $    0_{ -55}^{ +20}$  \\  
  J1012+2613 & $ -3600_{-1200}^{+3100}$ & $ -820_{-280}^{+700}$  \\  
  J1027+6050 & $  +400_{-1900}^{+1800}$ & $  +60_{-250}^{+240}$  \\  
  J1050+3456 & $  +210_{ -300}^{ +300}$ & $  +40_{ -55}^{ +55}$  \\  
  J1105+0414 & $  +550_{-2300}^{+2400}$ & $  +80_{-330}^{+350}$  \\  
  J1117+6741 & $ -2300_{-1000}^{ +900}$ & $ -260_{-110}^{+100}$  \\  
  J1154+0134 & $  -100_{-1800}^{ +400}$ & $  -15_{-275}^{ +55}$  \\  
  J1207+0604 & $ -1100_{ -800}^{ +700}$ & $ -130_{ -90}^{ +80}$  \\  
  J1211+4647 & $  +800_{ -370}^{ +790}$ & $ +130_{ -60}^{+130}$  \\  
  J1215+4146 & $  -150_{-2400}^{+1400}$ & $  -20_{-370}^{+210}$  \\  
  J1216+4159 & $ +1200_{-1400}^{+1000}$ & $ +190_{-220}^{+150}$  \\  
  J1328-0129 & $  -650_{ -370}^{ +310}$ & $  -80_{ -40}^{ +40}$  \\  
  J1414+1658 & $ -1800_{ -350}^{ +350}$ & $ -540_{-110}^{+110}$  \\  
  J1440+3319 & $ +2900_{ -750}^{ +730}$ & $ +500_{-130}^{+130}$  \\  
  J1536+0441 & $   +70_{ -290}^{ +230}$ & $  +20_{-110}^{ +80}$  \\  
  J1539+3333 & $  -190_{ -350}^{ +350}$ & $  -30_{ -55}^{ +55}$  \\  
  J1652+3123 & $  -100_{-2700}^{+3000}$ & $  -20_{-530}^{+590}$  \\  
  J1714+3327 & $   +50_{ -820}^{ +420}$ & $  +20_{-300}^{+150}$  \\  
\hline
 \end{tabular}
 \end{center}
\end{table}

\begin{figure*}
\includegraphics[width=0.99\columnwidth]{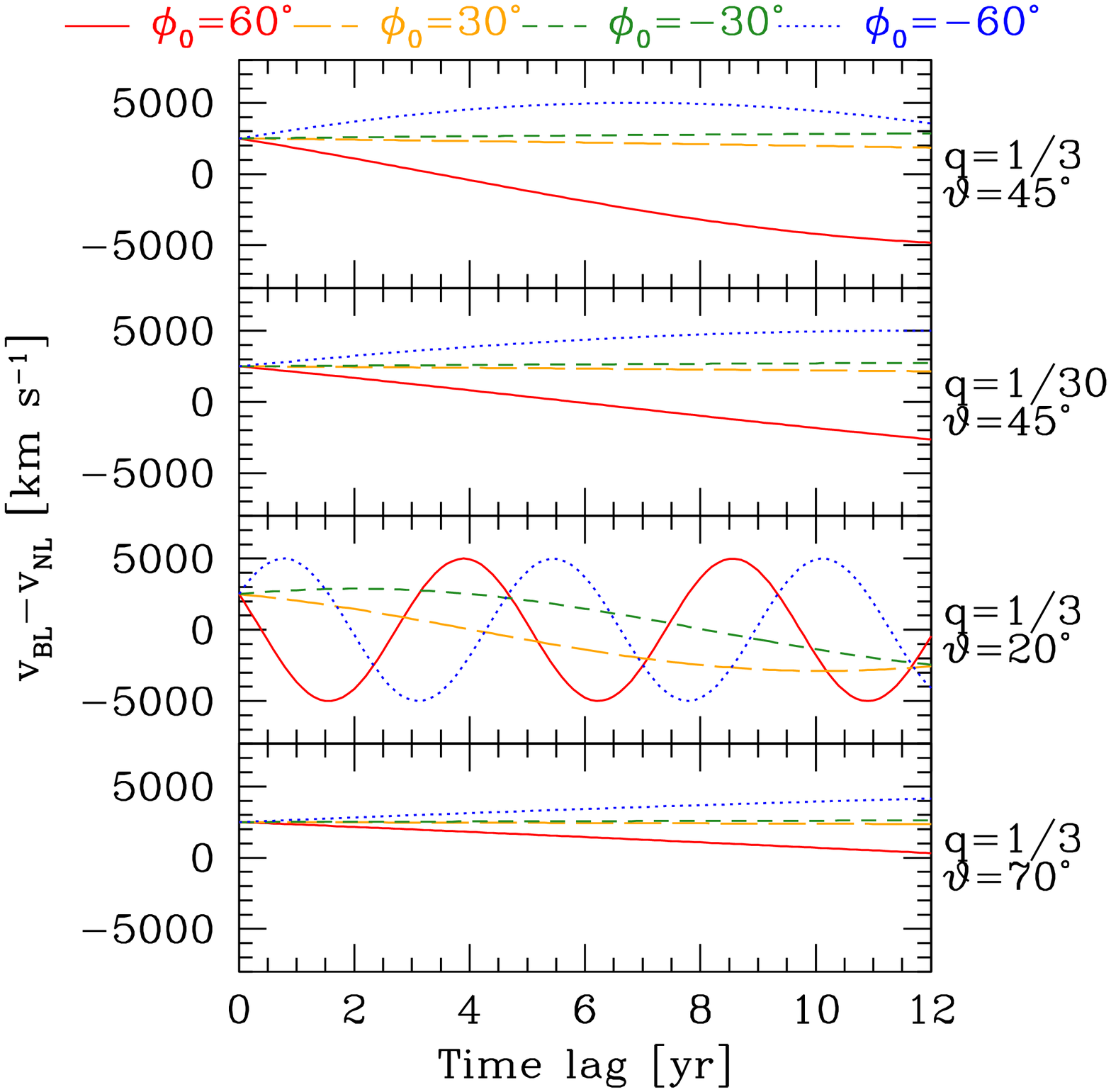}
\includegraphics[width=0.99\columnwidth]{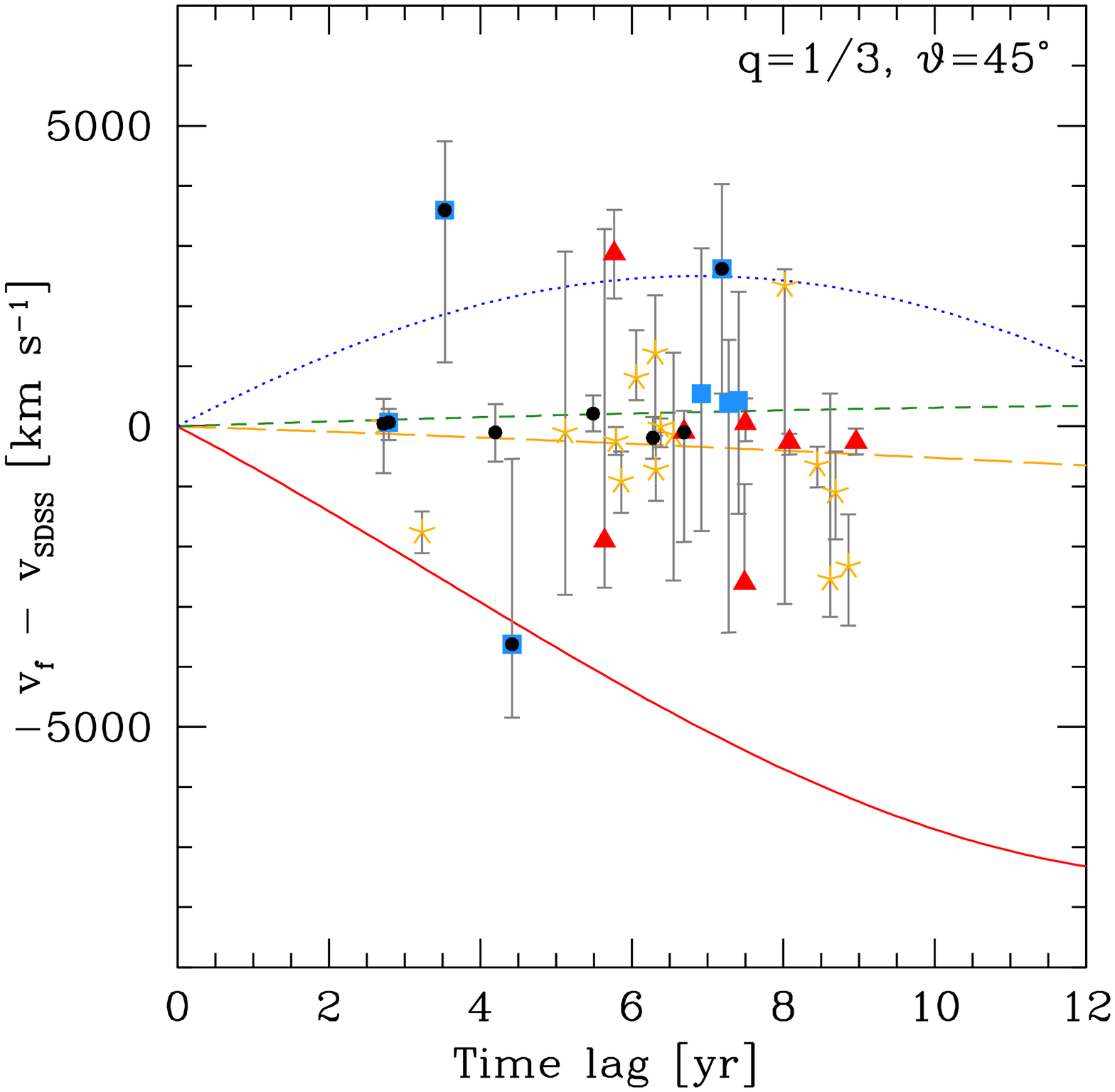}
\caption{{\em Left:} Expected evolution of the observed velocity shift for
a BHB with primary BH mass $M_1$=$10^9$ \Msun{}, initial velocity shift between 
BLs and NLs of 2500 \kms{}, and various values of the BH mass ratio
$q=M_2/M_1$, of the inclination angle of the orbit plane $\vartheta$, and
of the initial phase $\phi_0$. {\em Right:} The observed difference between
the BL peak shifts in our follow-up spectra ($v_{\rm f}$) and in SDSS data. 
The colour code highlights the source classification, as in Figure 
\ref{fig_linediag}. For a comparison,
we overplot the expected evolution predicted in the case of $M_1$=$10^9$ \Msun{},
$q=1/3$, $\vartheta=45^{\circ}$, assuming various orbital phases of the binary
at the time of the SDSS observations.}
\label{fig_velevo}
\end{figure*}
A key prediction of the BHB scenario is that the peaks of
BLs should shift periodically from longer to shorter wavelengths
(and vice versa) with respect to NLs, over orbital periods which 
are estimated to be 1--500 yr for the most reasonable geometrical 
configurations and BH masses 
\citep[see Figure \ref{fig_velevo}, {\em left}, and][]{dotti09,lauer09}.

Variations in the profiles of BLs have been observed in several AGN.
A clear-cut example is the evolution of the \Hb{} line profile in NGC
5548 over 30 yr \citep[][]{sergeev07}\footnote{For an animation of the
  line profile evolution, see
  \textsf{http://www.astronomy.ohio-state.edu/$\sim$peterson/AGN/30year.avi}},
where blue- and red-shifted peaks or double-horned profiles are
observed for the same object in different epochs. Usually these features
have modest entity (e.g., peak shifts $\Delta v\lsim 500$ \kms{}), evolve 
irregularly and dissipate out in less than an orbital period of the BL
region. The generally accepted picture is that they arise in dynamically 
unstable `bright spots' of the BL region. 

Monitoring the evolution of BL profiles over 5--10 years (i.e., a 
significant fraction of the BL region orbital period) is required
in order to distinguish between the BHB hypothesis and other scenarios. 
If the shifted peaks of BLs change their fluxes or shapes, without 
any predictable behavior (in particular, with no periodic behaviour),
then the `bright-spot' scenario is favoured. On the contrary, if the
peak shifts periodically around the NL redshift over several periods, 
then the BHB interpretation would be confirmed. 

In order to quantify the evolution in the line profiles, we first
fit the peak of each BL with a single gaussian. Compared to the
fitting method described in Section \ref{sec_spectra}, this approach 
performs better in identifying the peak wavelength in the case of
steep profiles (as observed in a number of our sources), whereas the
Gauss-Hermite approach provides a better description of the overall
line properties. We perform this analysis on both our follow-up data
and the SDSS spectra. Our measurements of the peak velocity differences, 
their uncertainties, and the corresponding mean accelerations are
reported in Table \ref{tab_evo} and plotted in Figure \ref{fig_velevo},
{\em right}. Velocity shifts are computed on the \Ha{} profiles, unless
only \Hb{} is available. Prominent ($>$ 2$\sigma$) changes in the BL peak wavelengths
are reported in 5 sources: J1117+6741, J1211+4647, J1328-0129, J1414+1658
(all classified as `Others') and J1440+3319 (`Asymmetric'). Less significant
changes are also reported for the `Asymmetric' sources J0012-1022 and 
J0936+5331, and for J1207+0604 (classified as `Others'). 

\begin{figure*}
\includegraphics[width=0.3\textwidth]{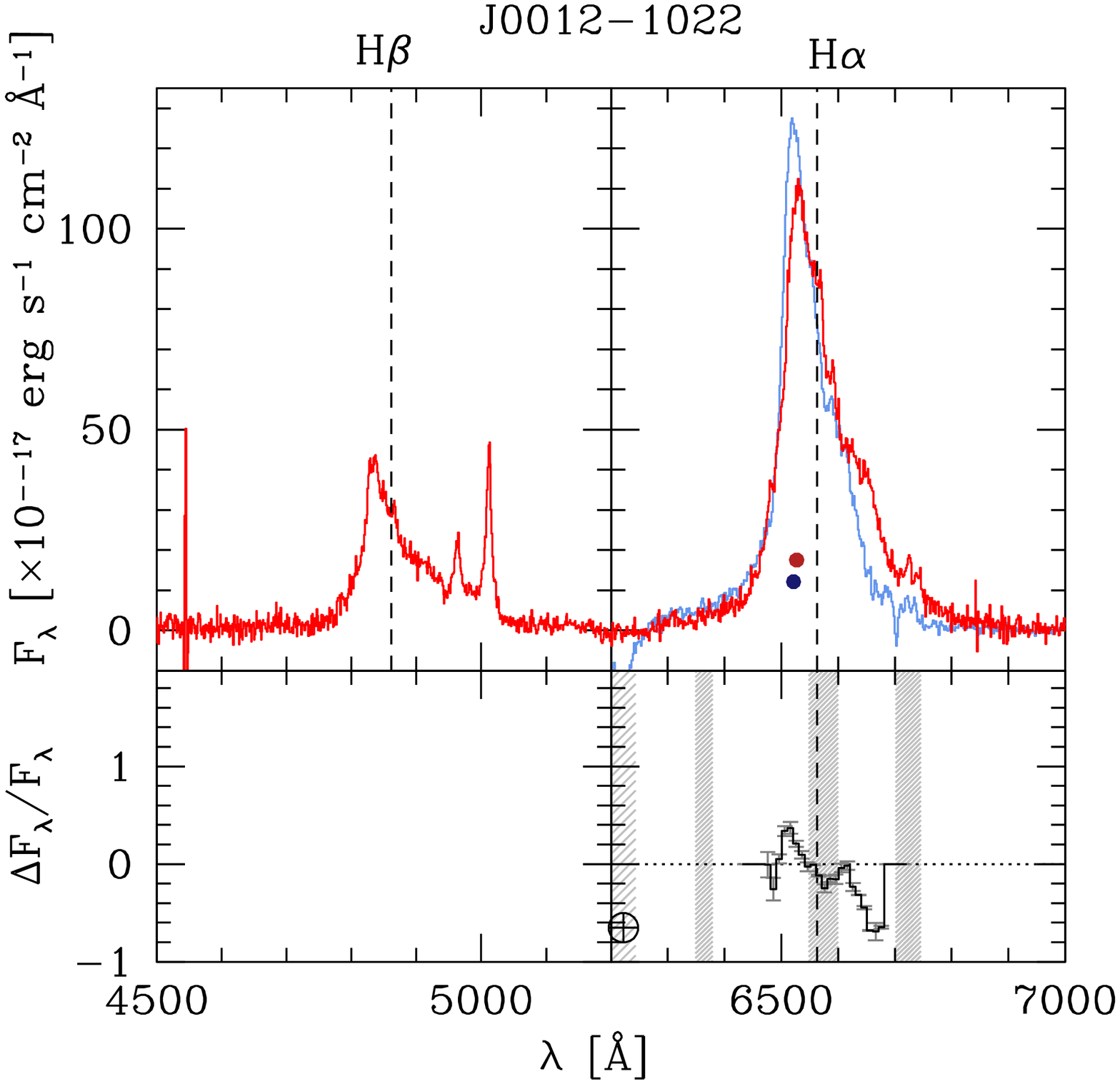} 
\includegraphics[width=0.3\textwidth]{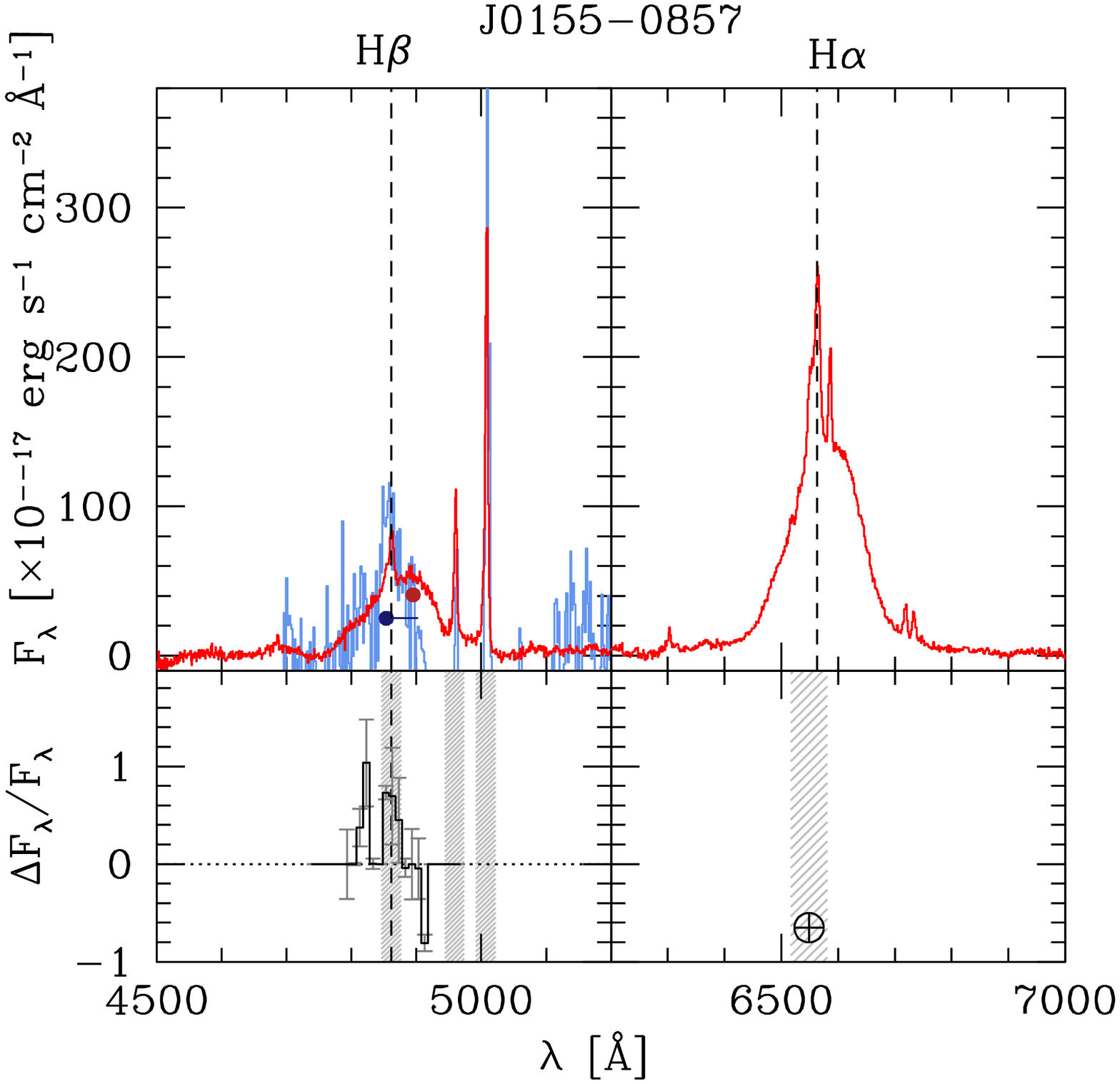} 
\includegraphics[width=0.3\textwidth]{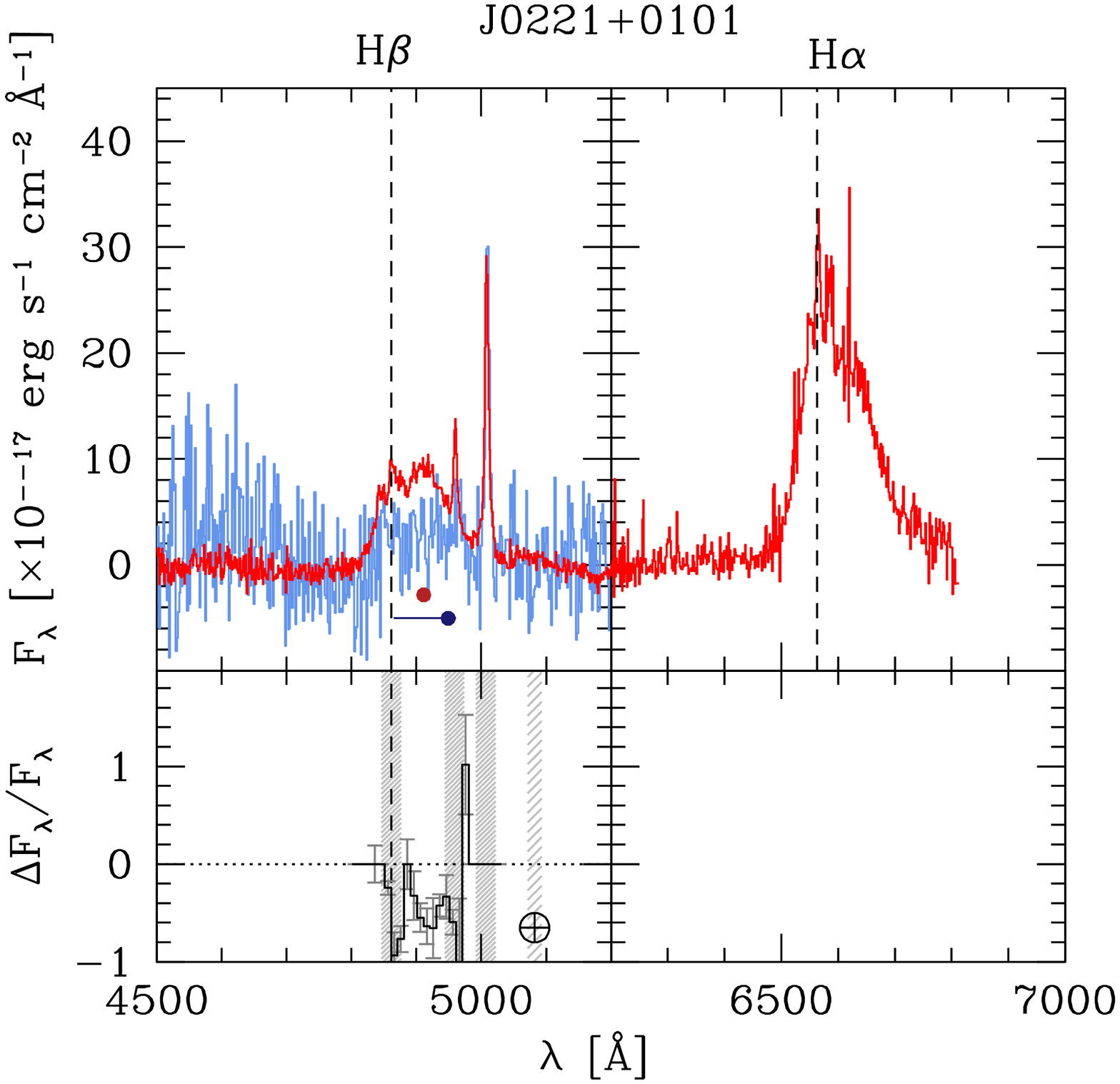} \\
\includegraphics[width=0.3\textwidth]{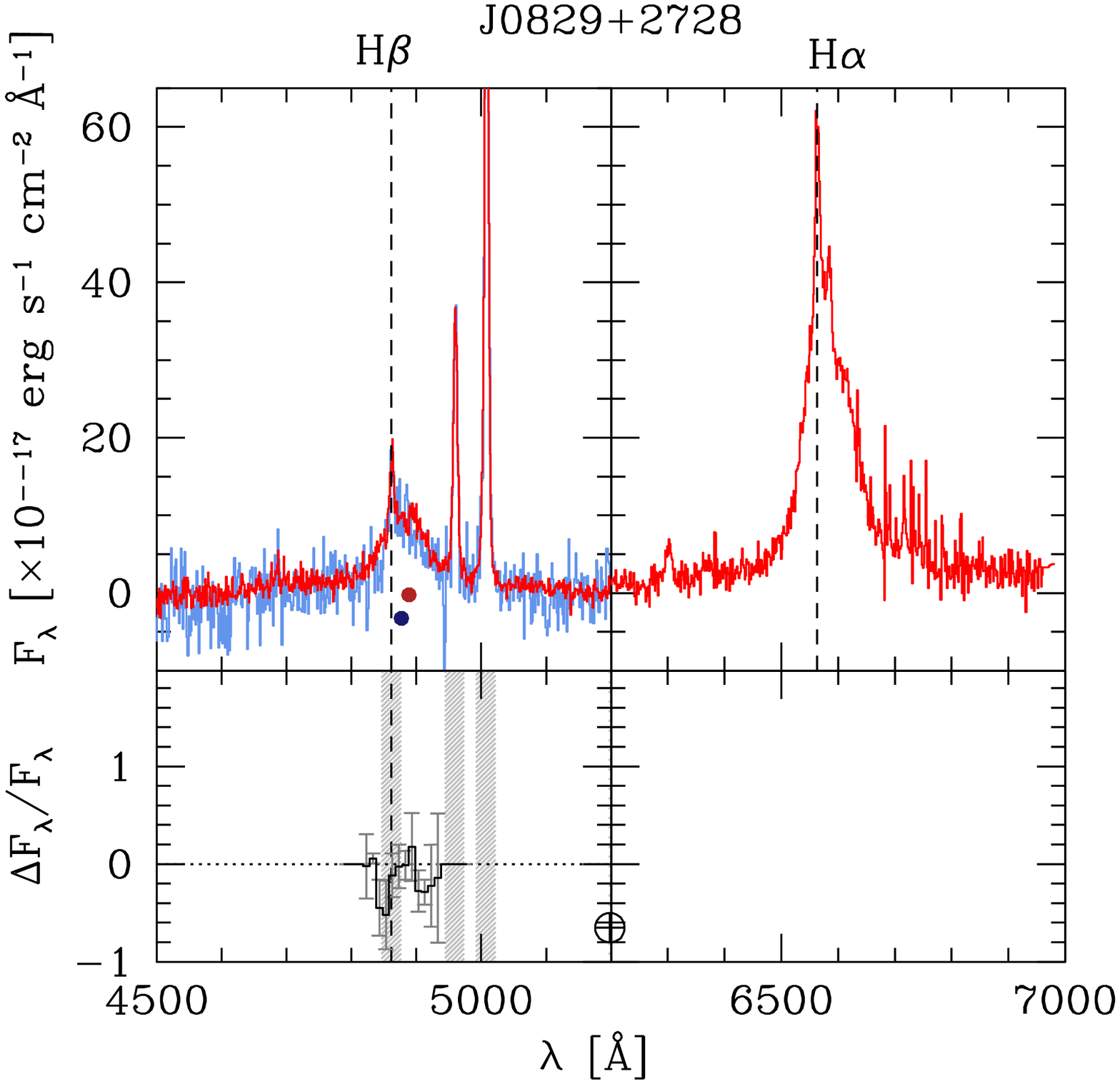} 
\includegraphics[width=0.3\textwidth]{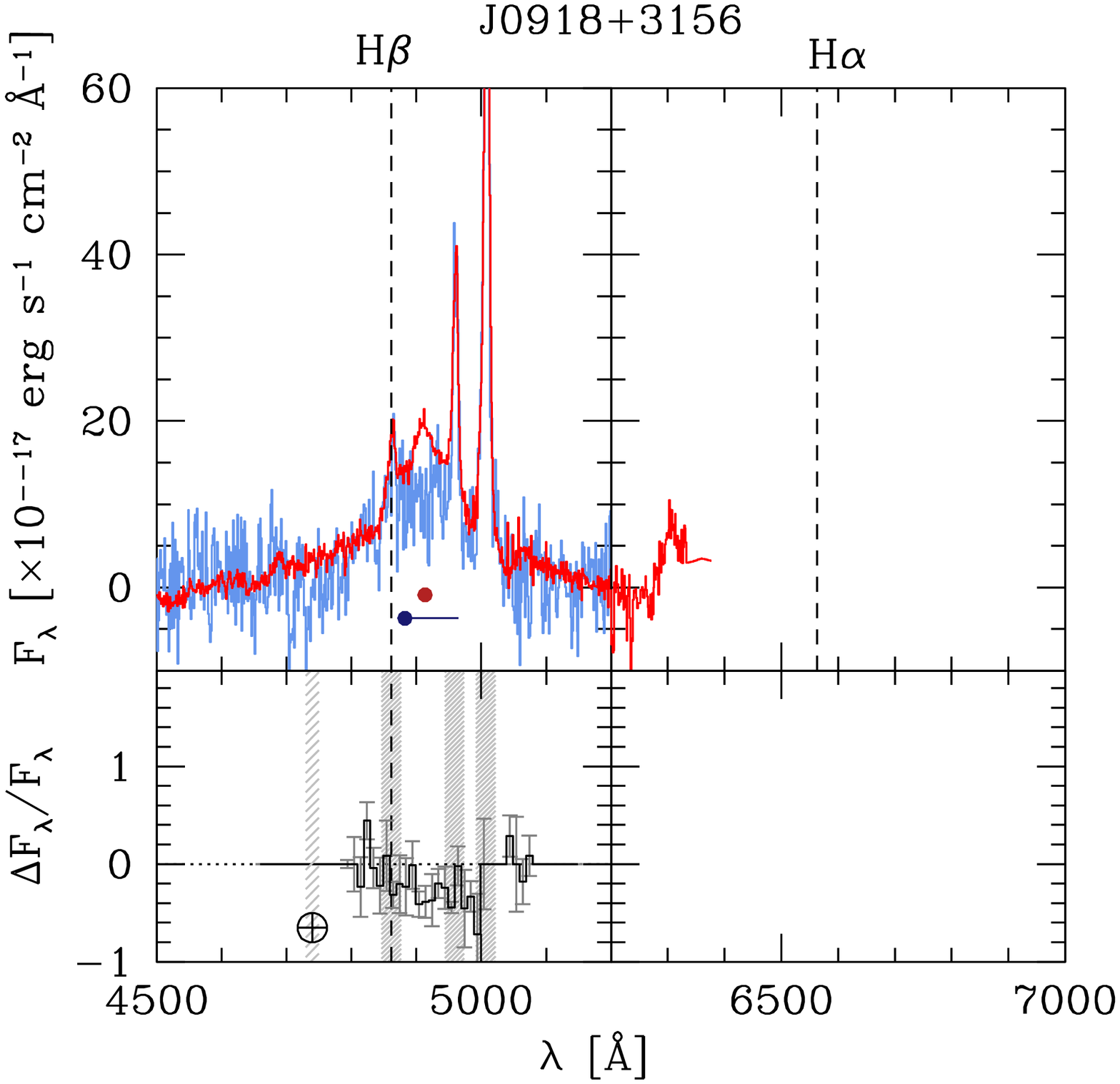} 
\includegraphics[width=0.3\textwidth]{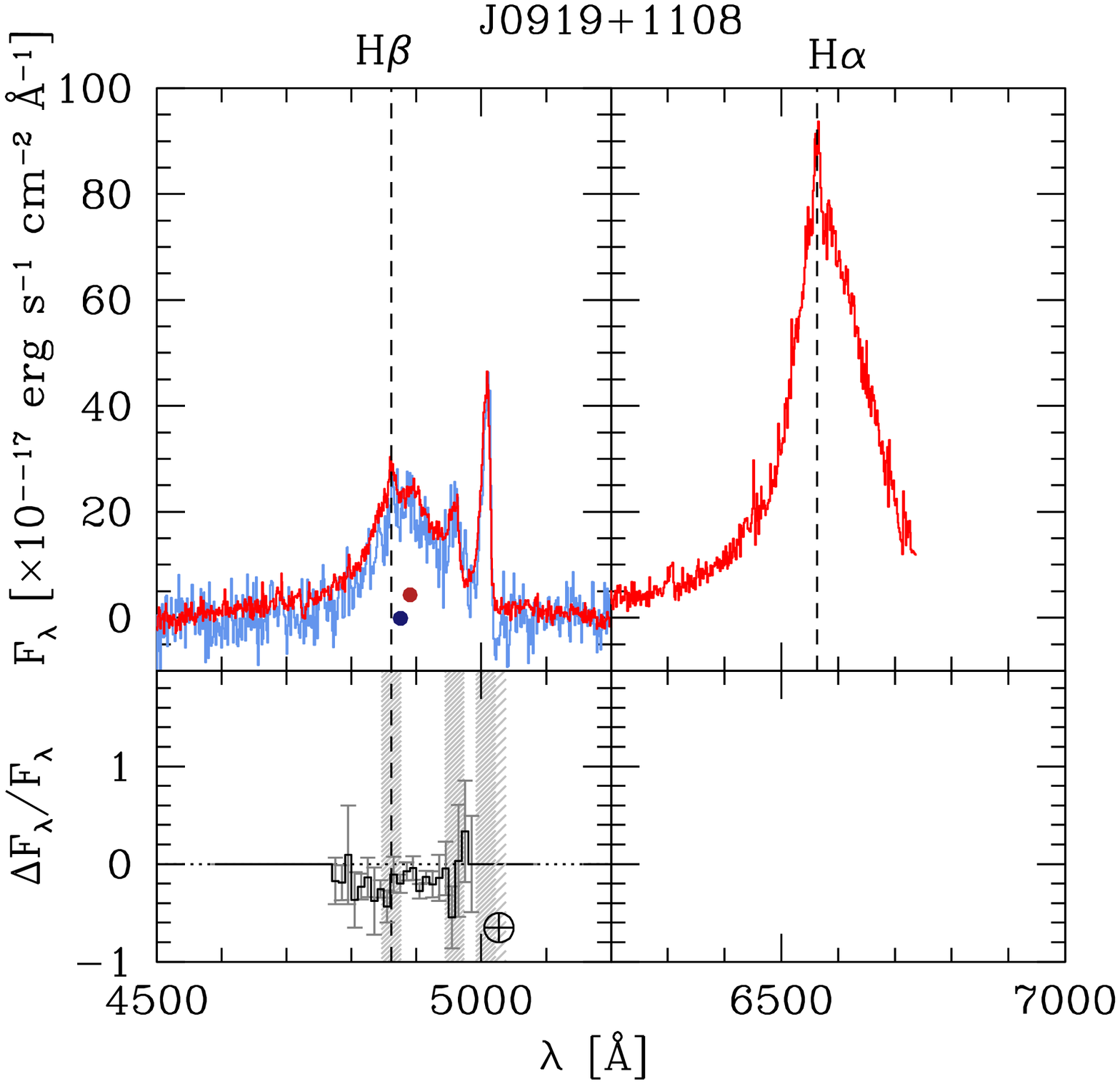} \\
\includegraphics[width=0.3\textwidth]{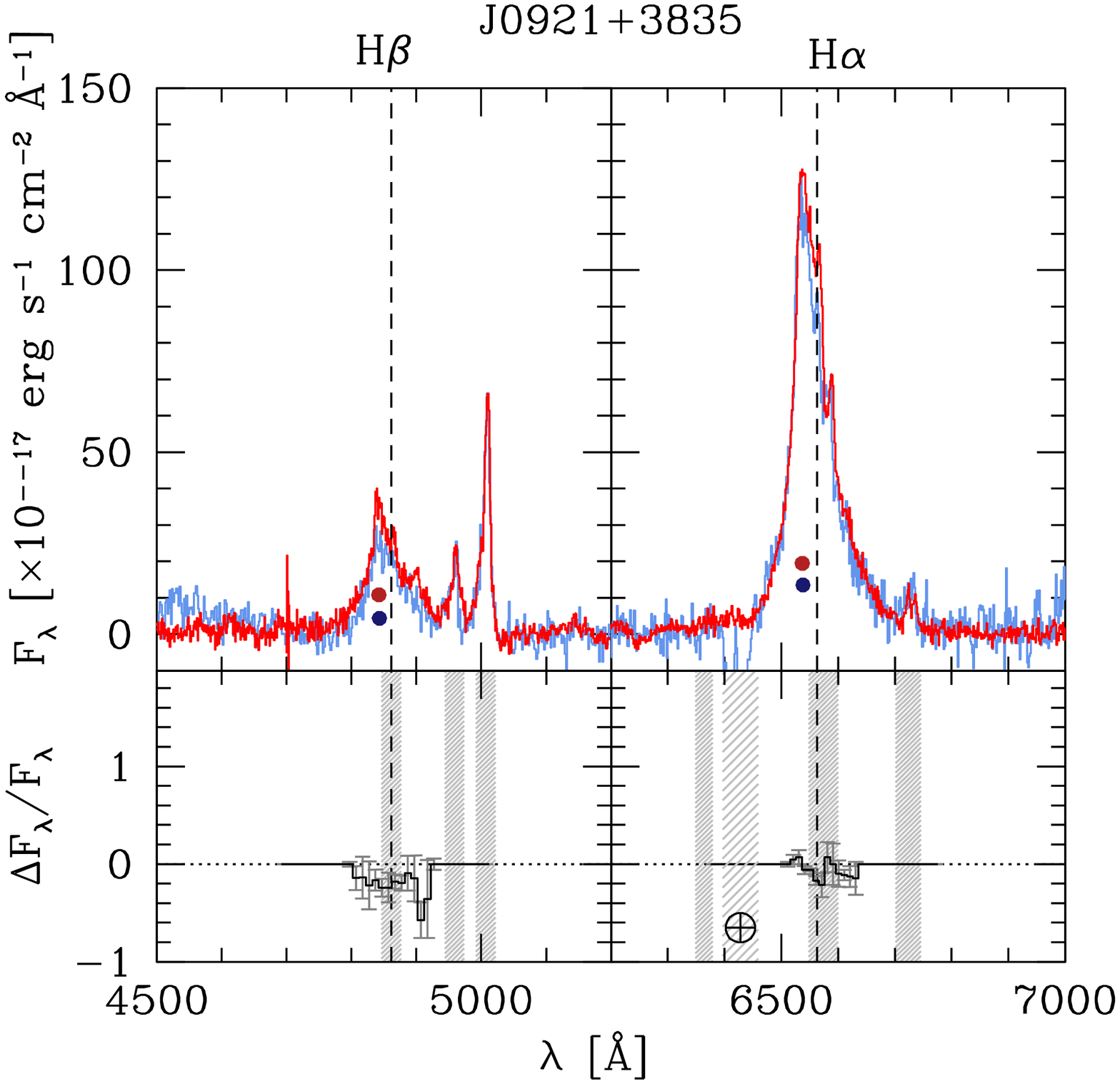} 
\includegraphics[width=0.3\textwidth]{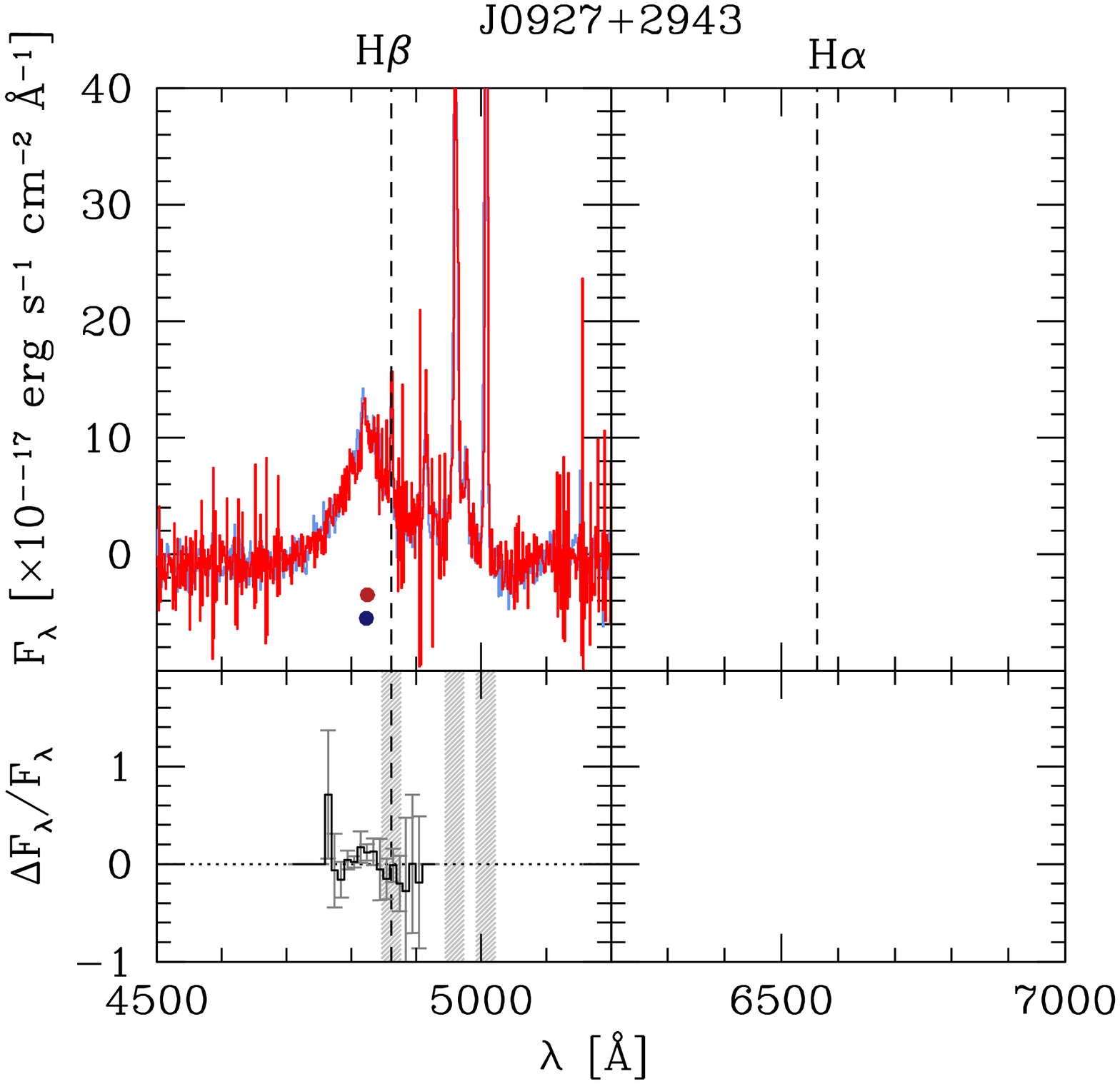} 
\includegraphics[width=0.3\textwidth]{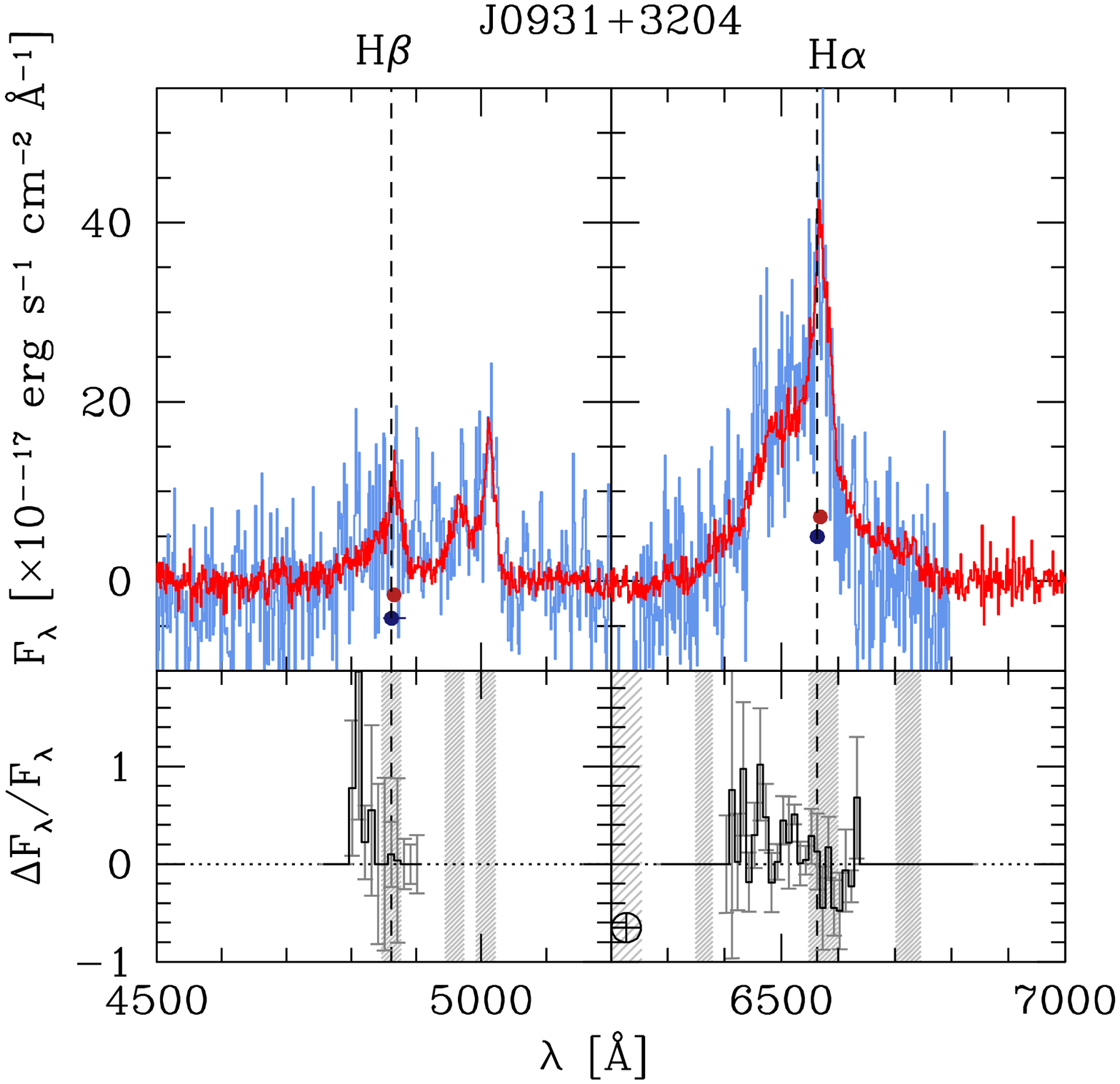} \\
\includegraphics[width=0.3\textwidth]{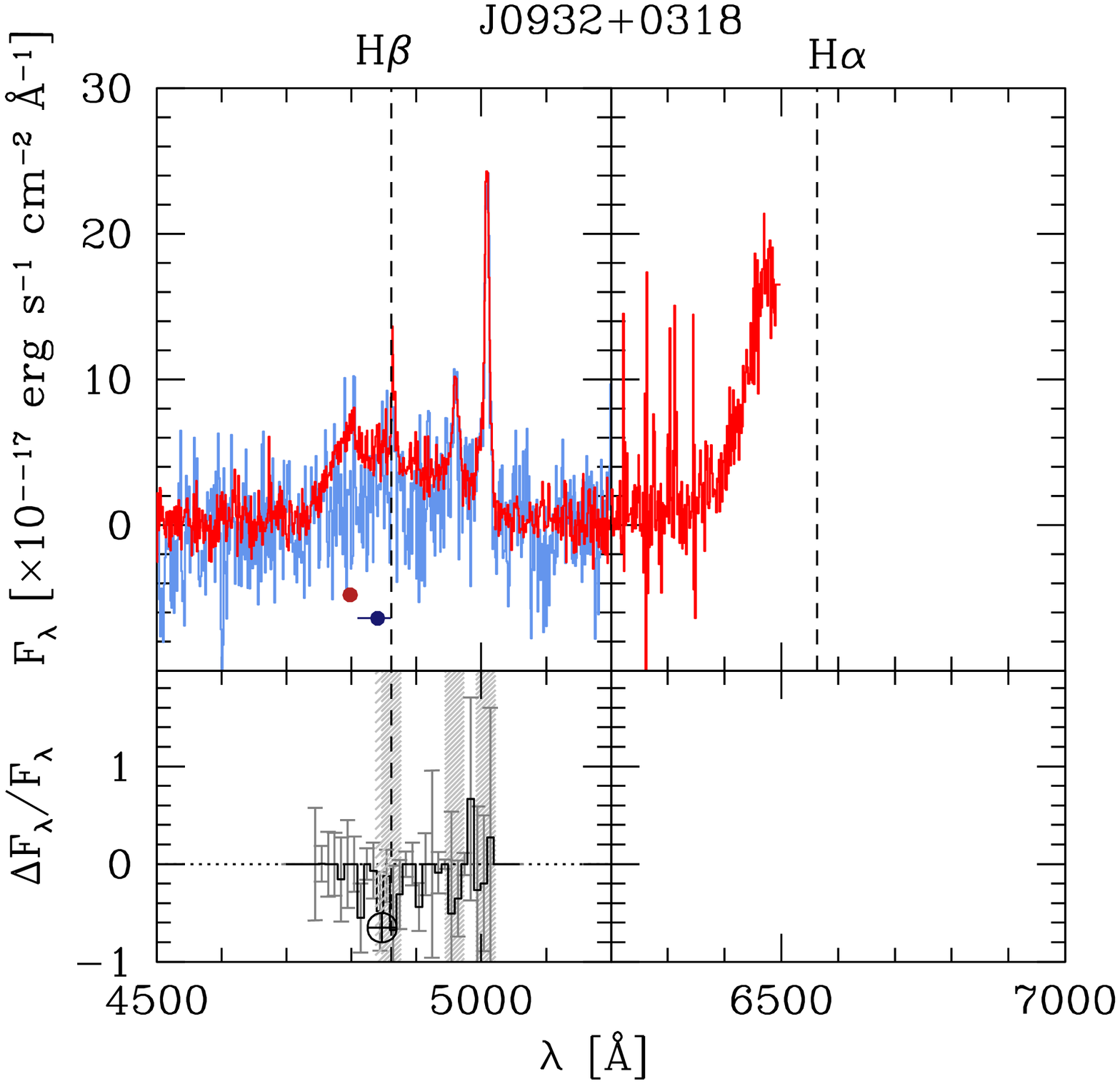} 
\includegraphics[width=0.3\textwidth]{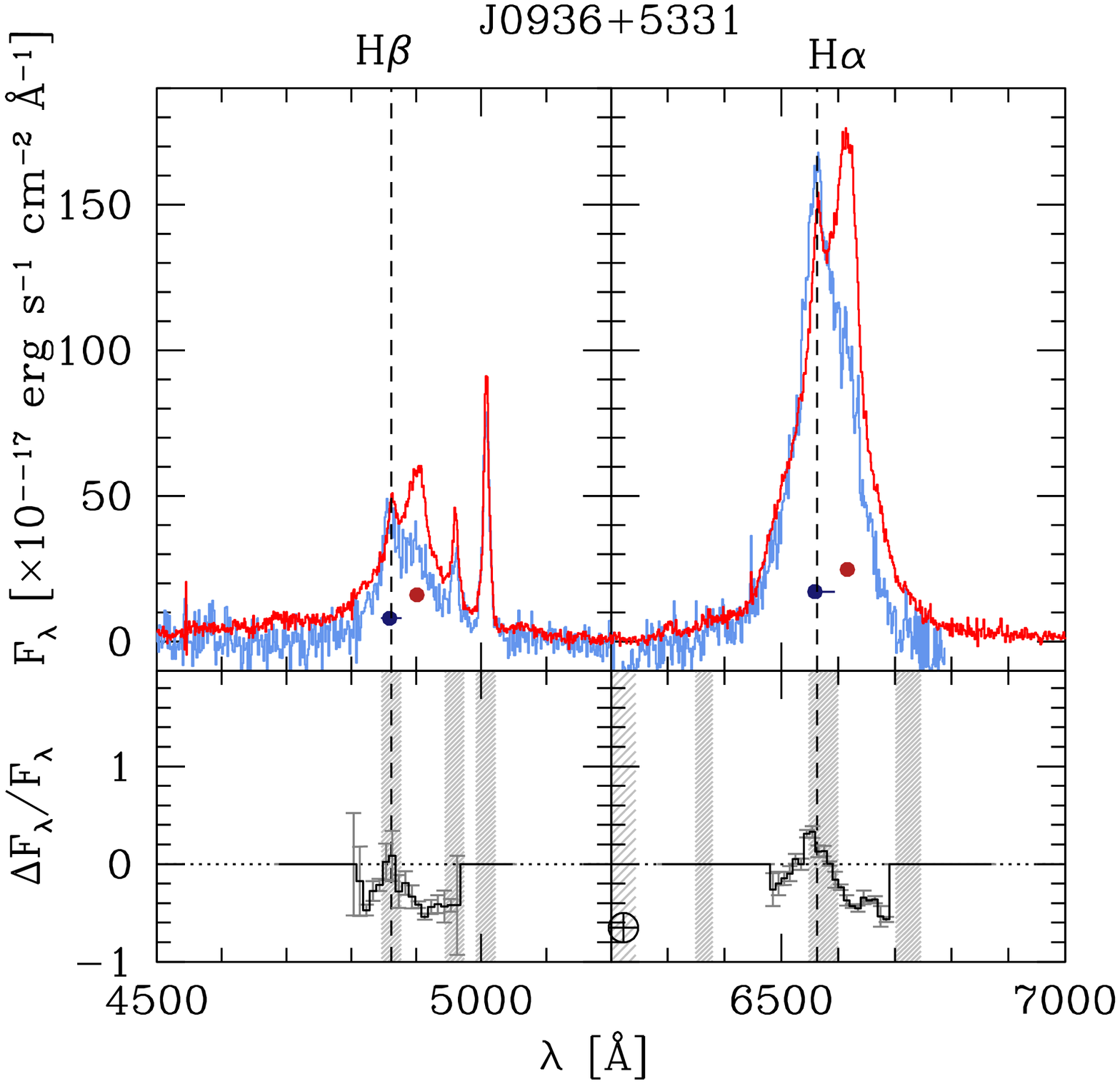} 
\includegraphics[width=0.3\textwidth]{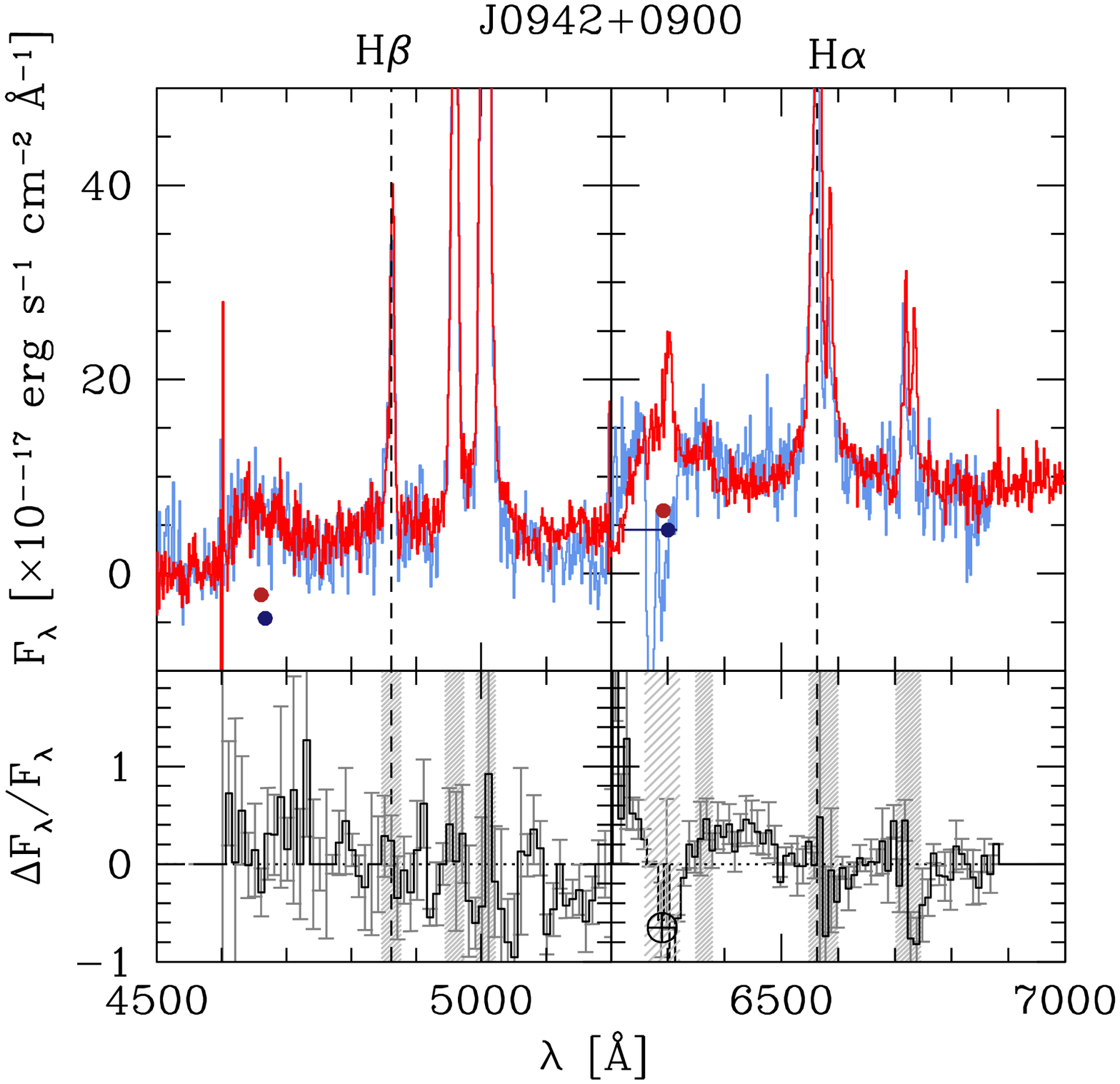} \\
\caption{Comparison between the SDSS data and our follow-up
  observations for the quasars in our sample. The {\em top panels} of
  each window show the \Hb{} (left) and \Ha{} (right) line profiles in
  the SDSS spectrum (red) and in our follow-ups
  (blue). Spectra have been continuum subtracted, and
  scaled so that the \Oiii{} and \Nii{} lines in the follow-up
  observations match the flux observed in the SDSS spectra (see text
  for details). The expected wavelengths of the narrow components of 
  \Ha{} and \Hb{} are highlighted with dashed, vertical lines, while
  the observed BL peak wavelengths are marked with dark red and blue points. 
  The {\em bottom panels} show the difference between the two epochs,
  normalized to the BL fit of the SDSS data and resampled in 10 \AA{} wide
  bins for the sake of clarity. Grey shaded area mark the position of
  prominent narrow emission lines and of atmospheric absorption features.}
\label{fig_spc1}
\end{figure*}

\begin{figure*}
\includegraphics[width=0.3\textwidth]{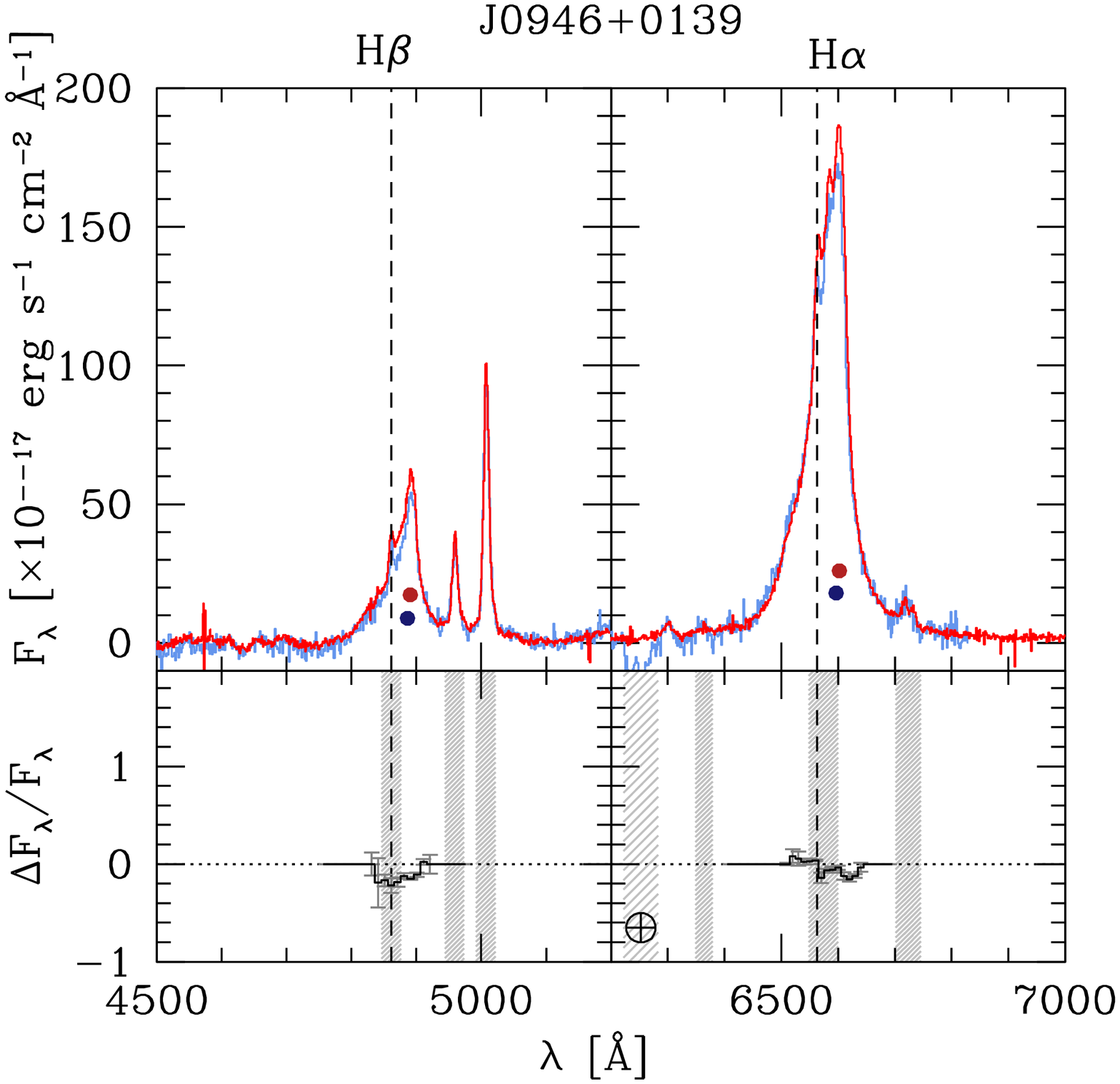} 
\includegraphics[width=0.3\textwidth]{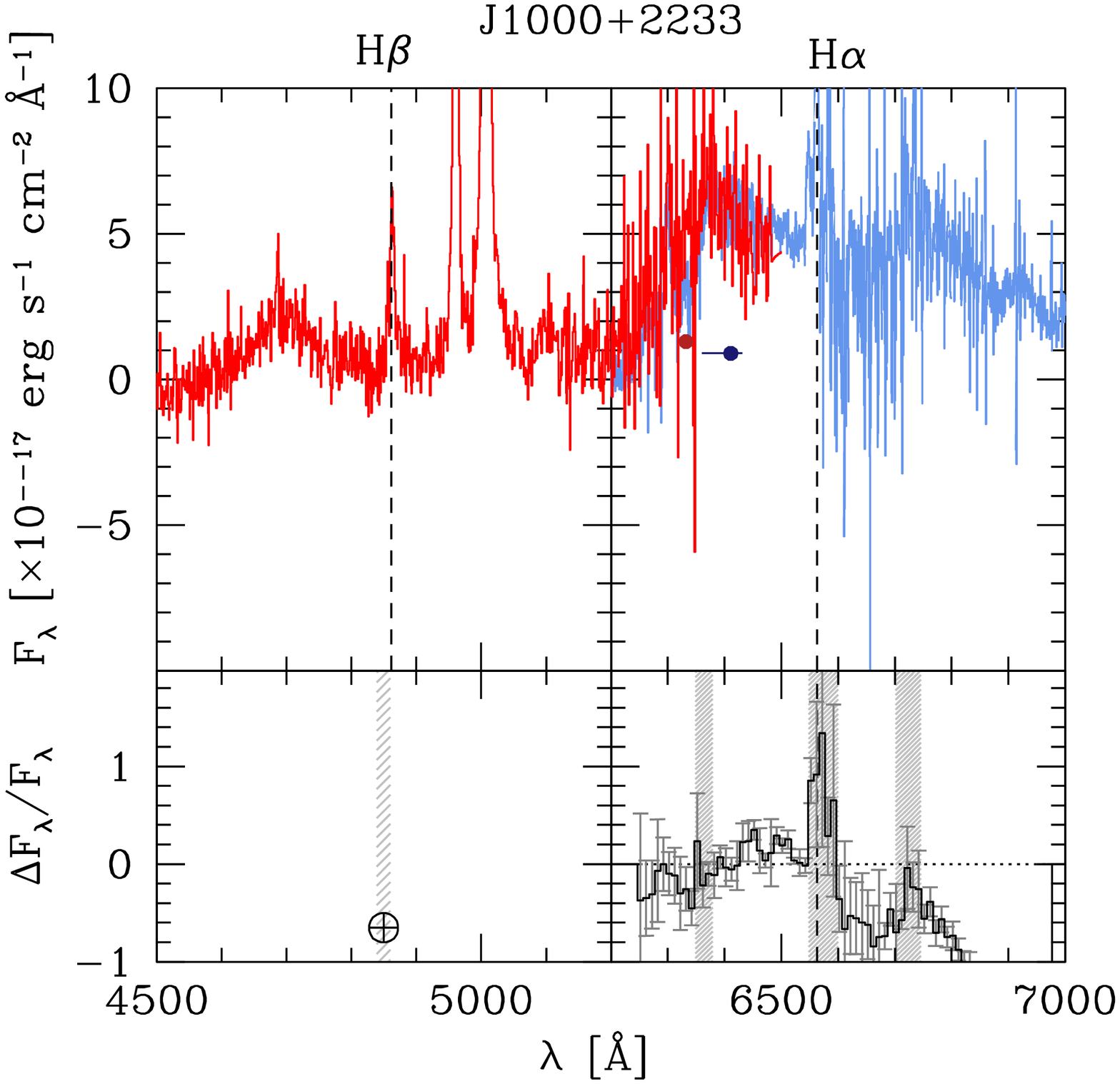} 
\includegraphics[width=0.3\textwidth]{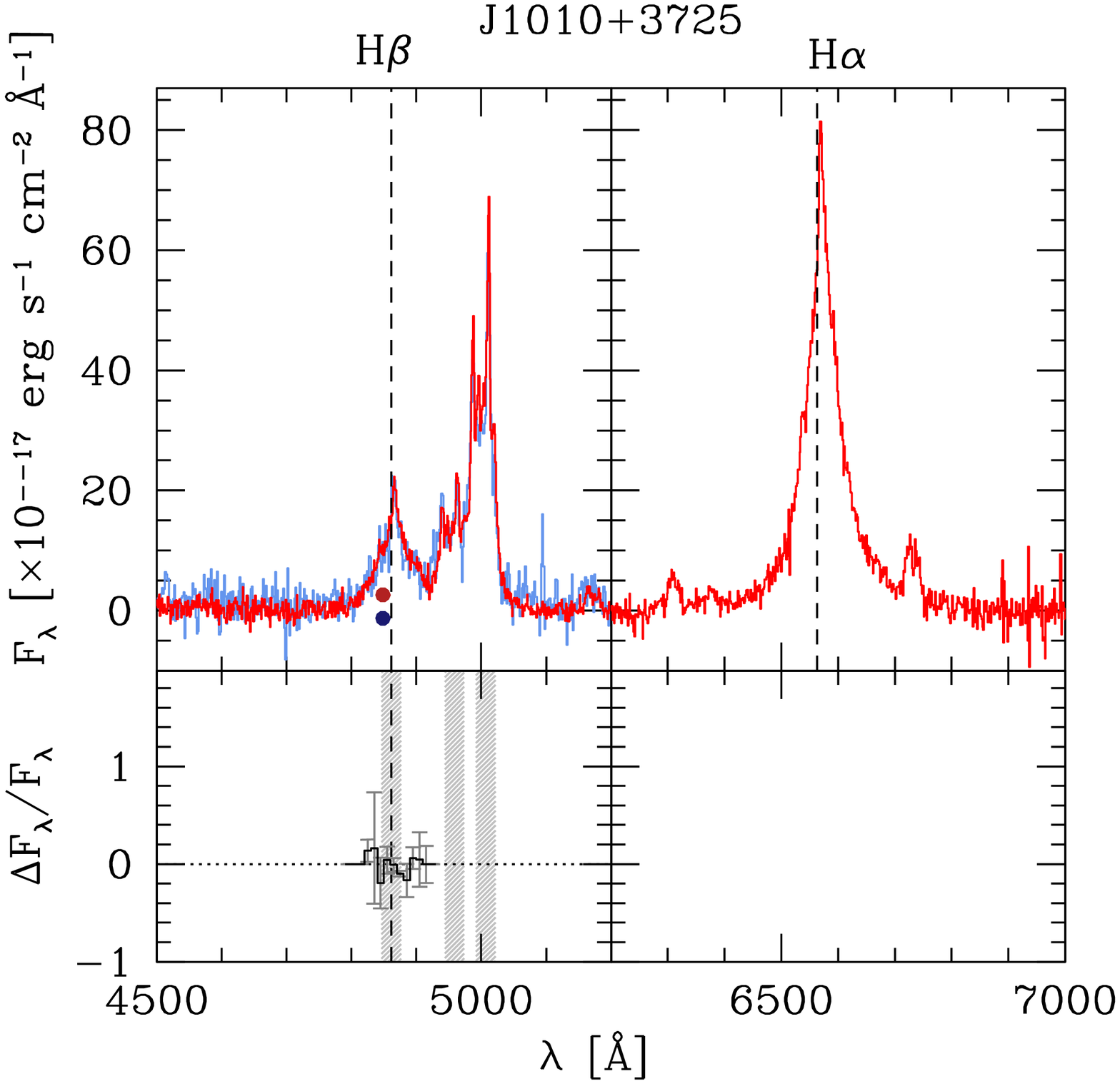} \\
\includegraphics[width=0.3\textwidth]{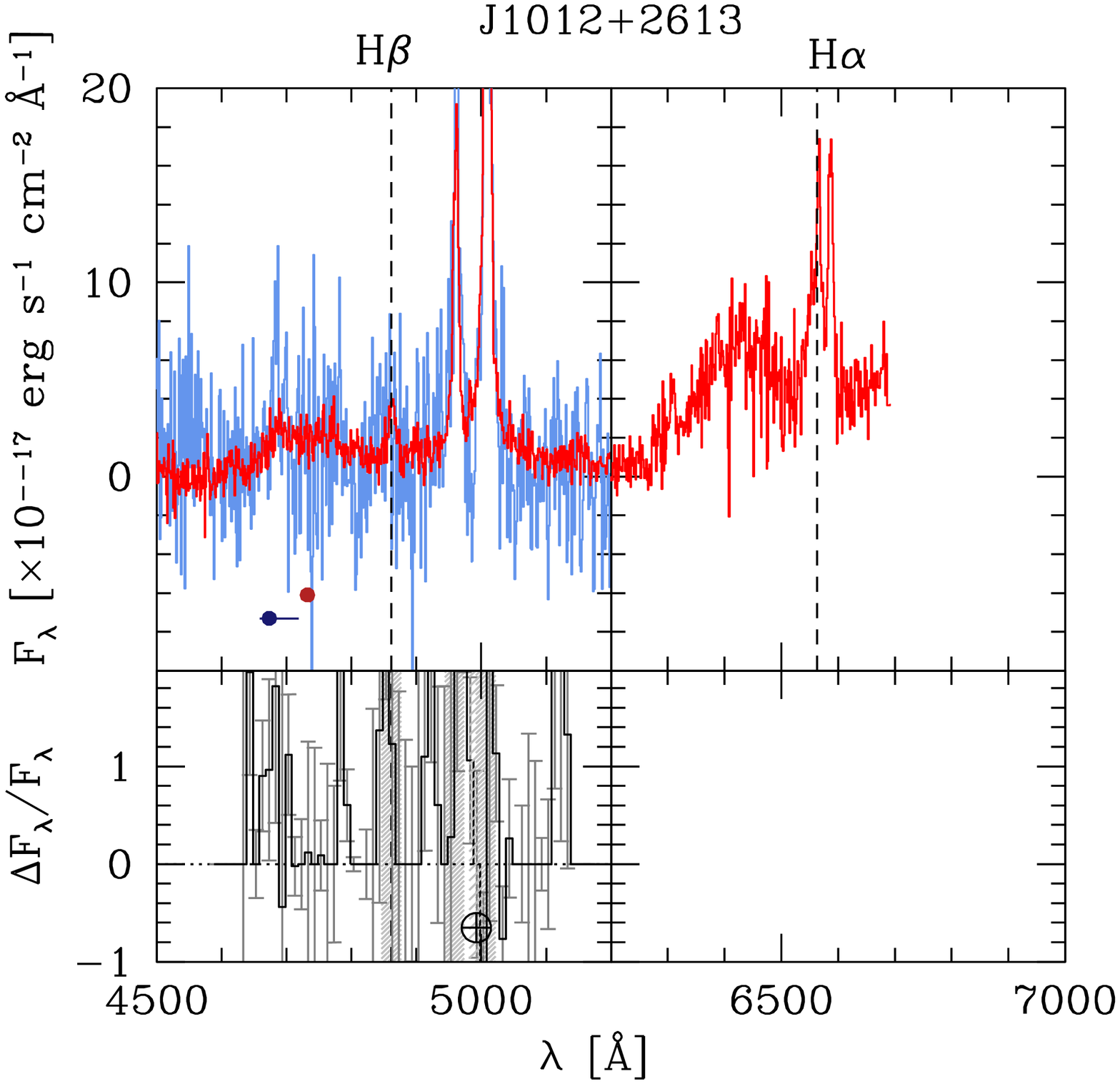} 
\includegraphics[width=0.3\textwidth]{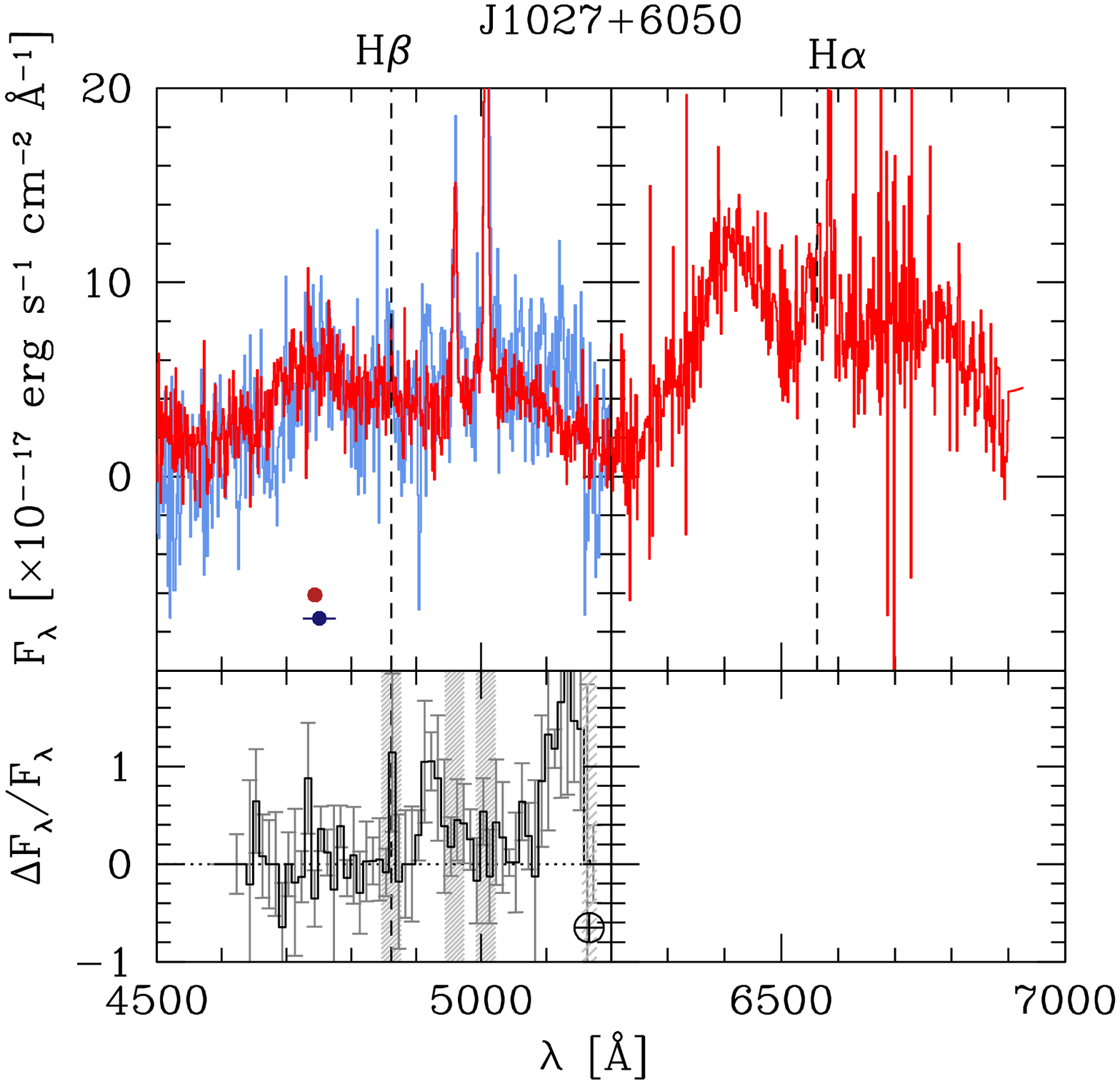} 
\includegraphics[width=0.3\textwidth]{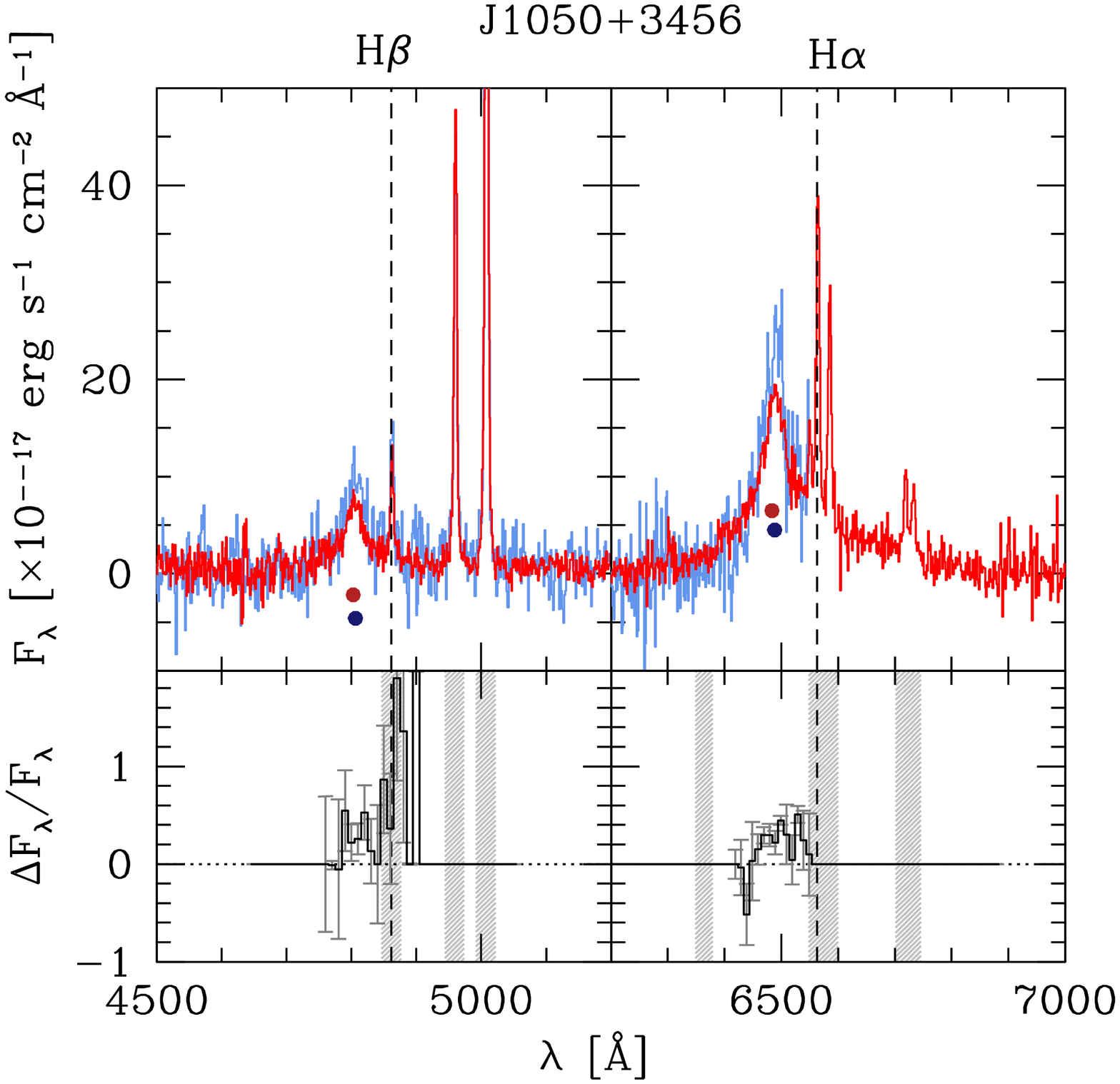} \\
\includegraphics[width=0.3\textwidth]{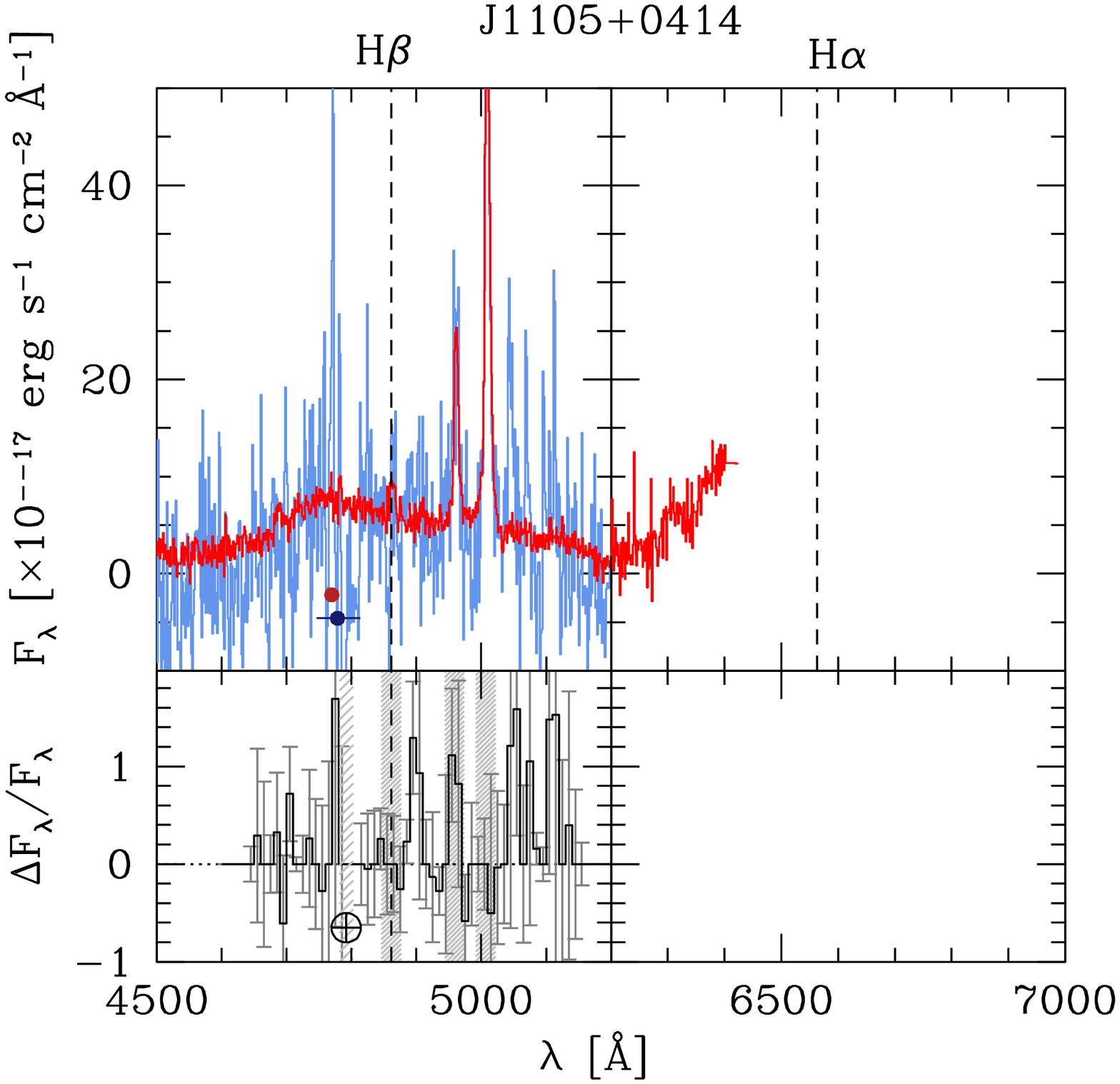} 
\includegraphics[width=0.3\textwidth]{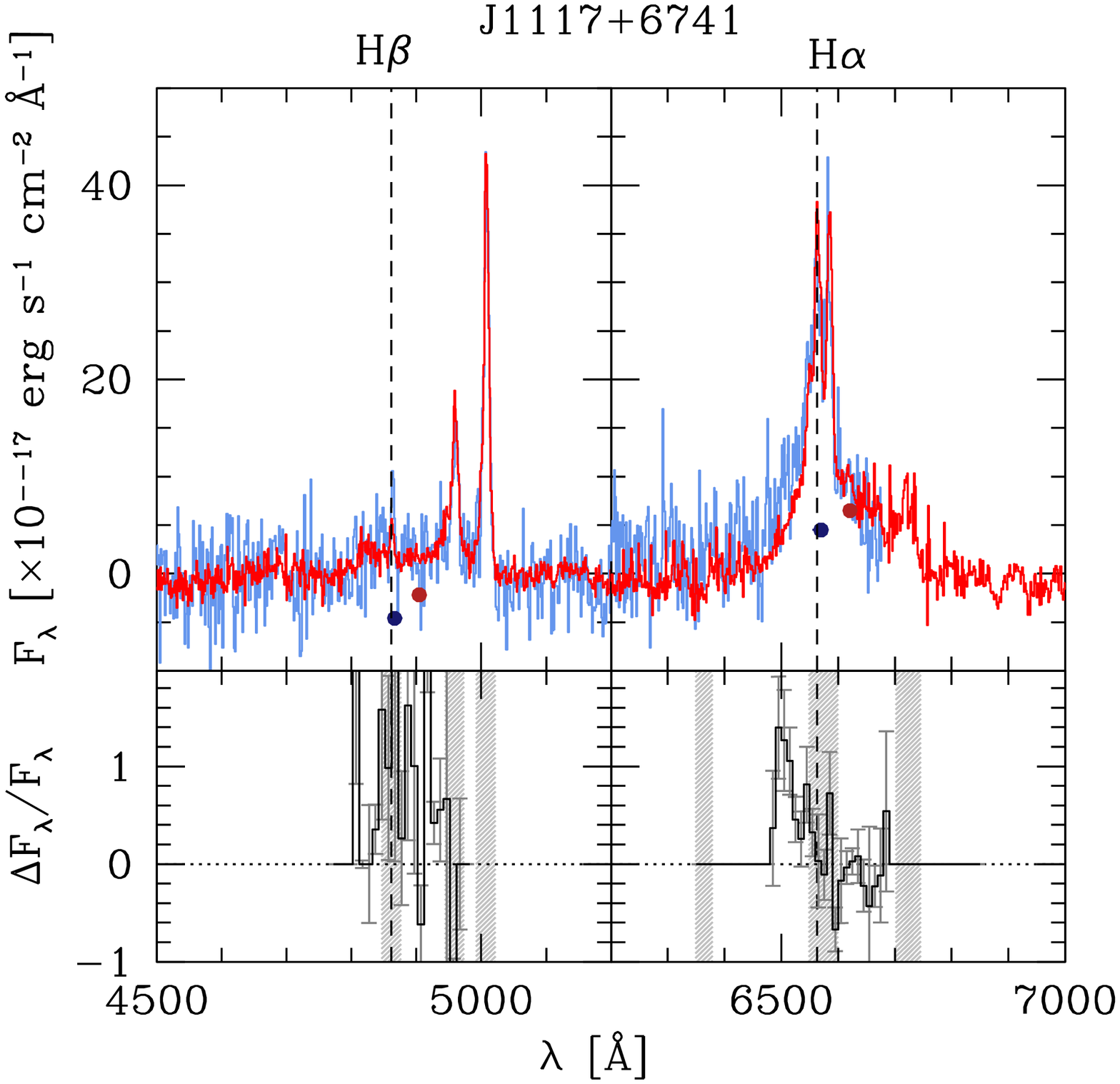} 
\includegraphics[width=0.3\textwidth]{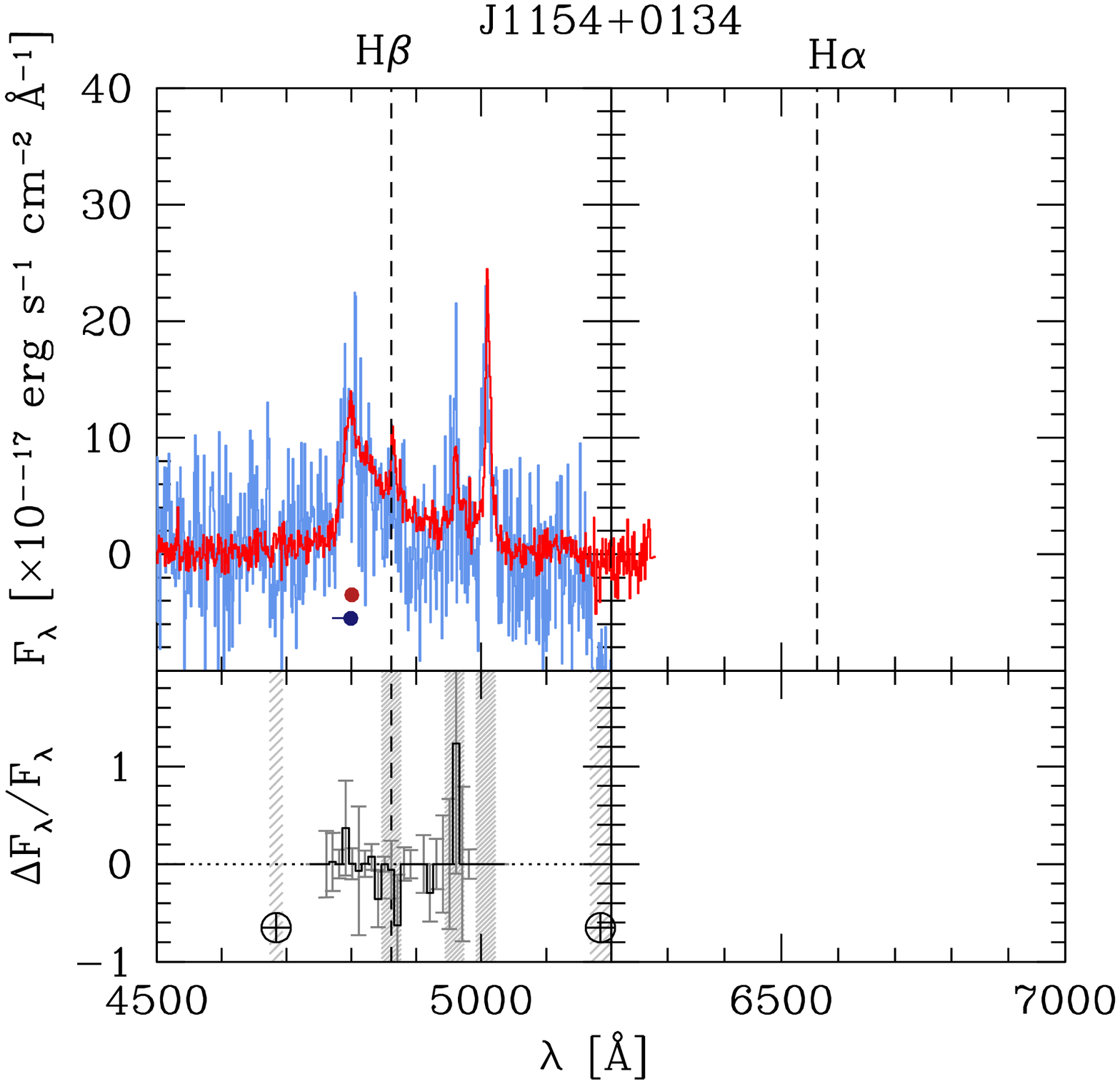} \\
\includegraphics[width=0.3\textwidth]{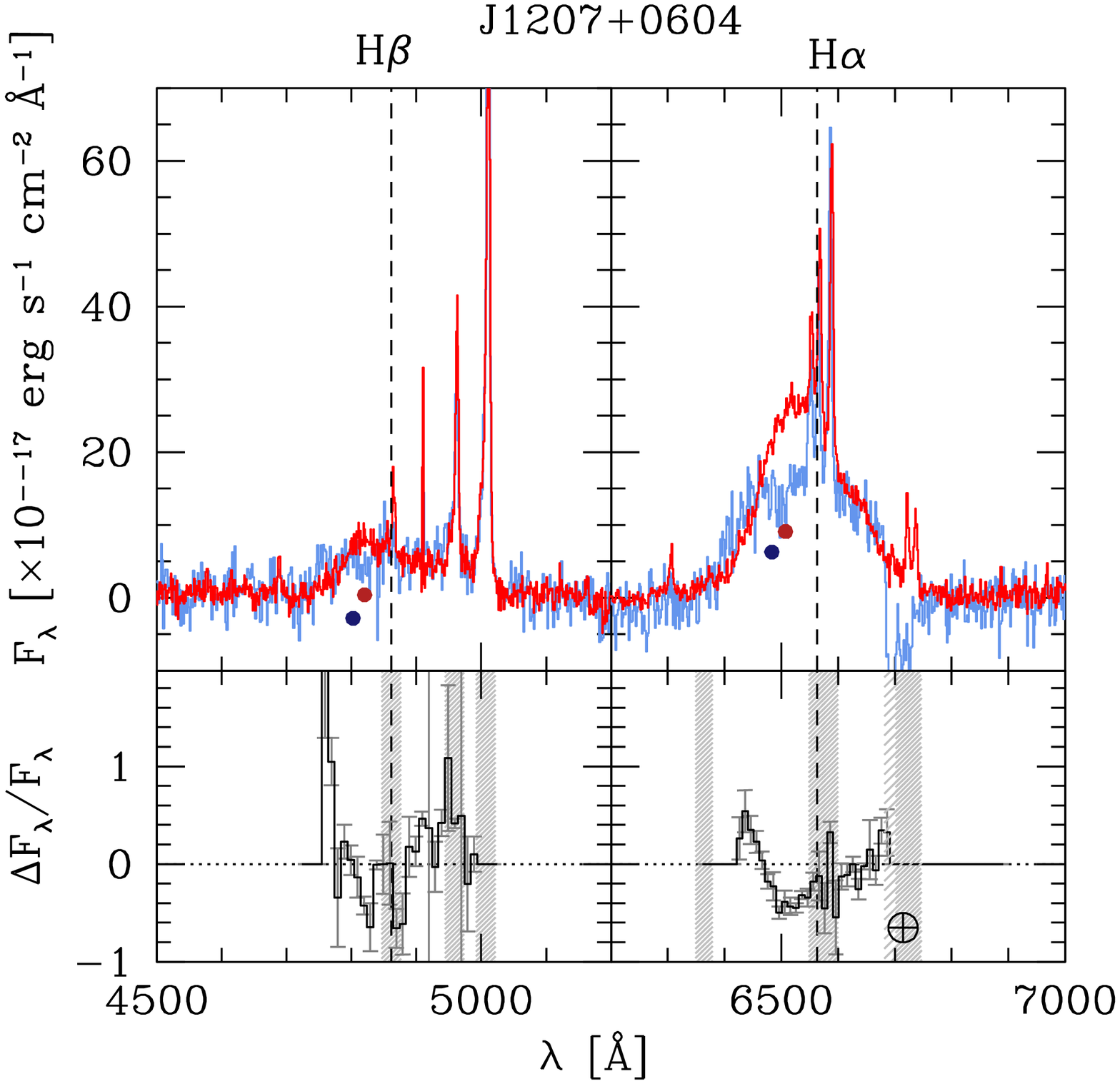} 
\includegraphics[width=0.3\textwidth]{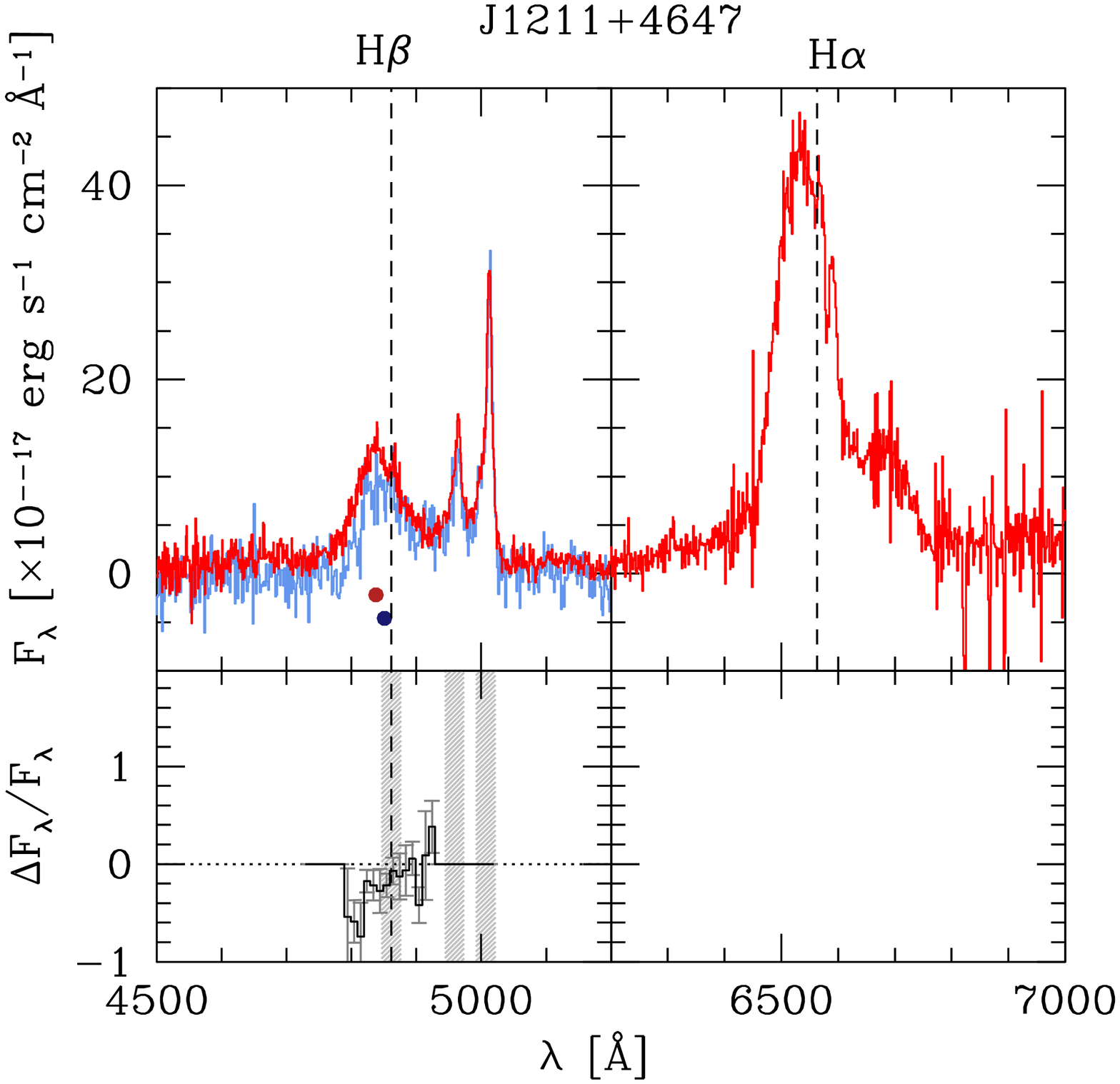} 
\includegraphics[width=0.3\textwidth]{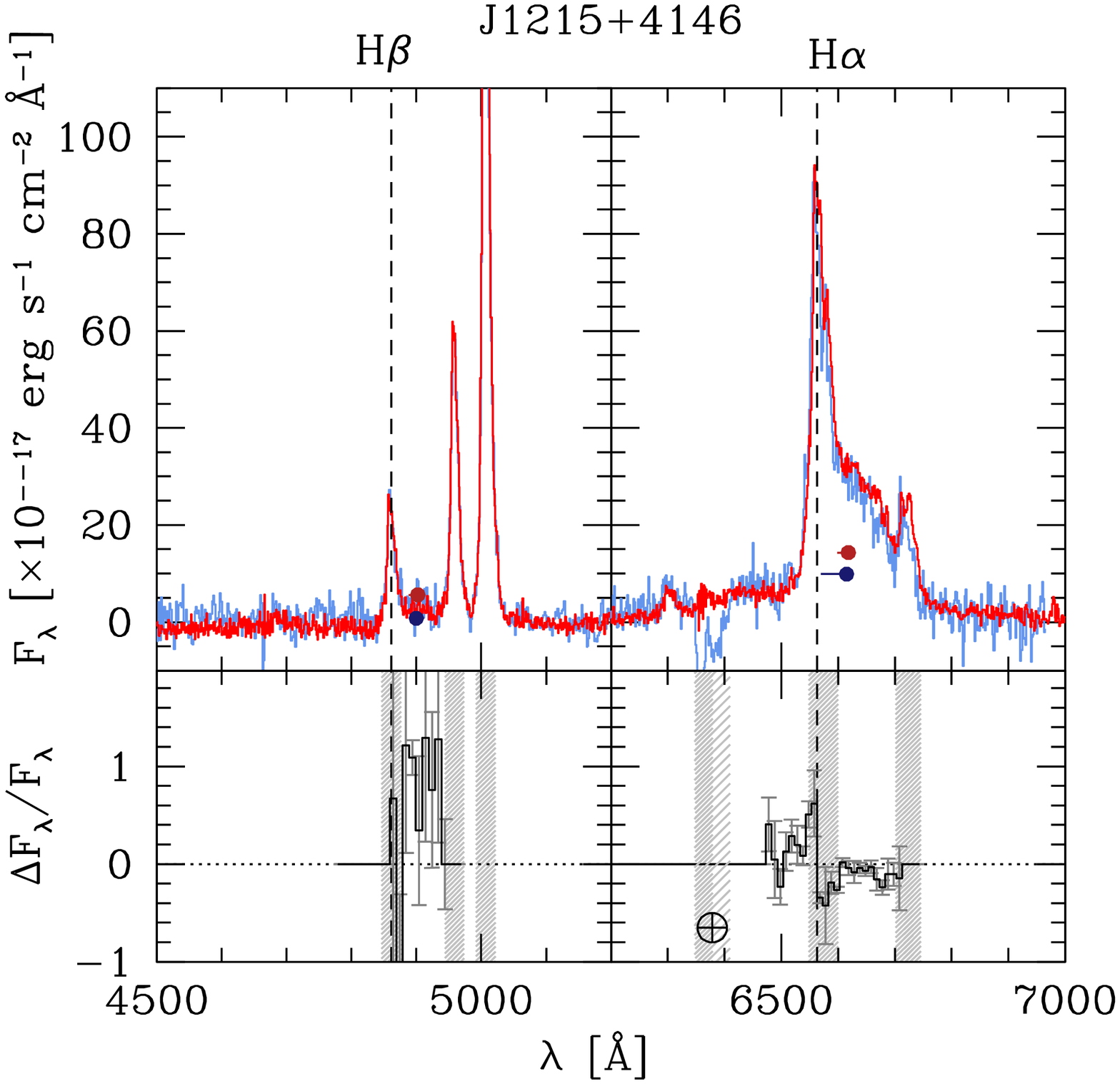} \\
\caption{Continue from Figure \ref{fig_spc1}.}
\label{fig_spc2}
\end{figure*}

\begin{figure*}
\includegraphics[width=0.3\textwidth]{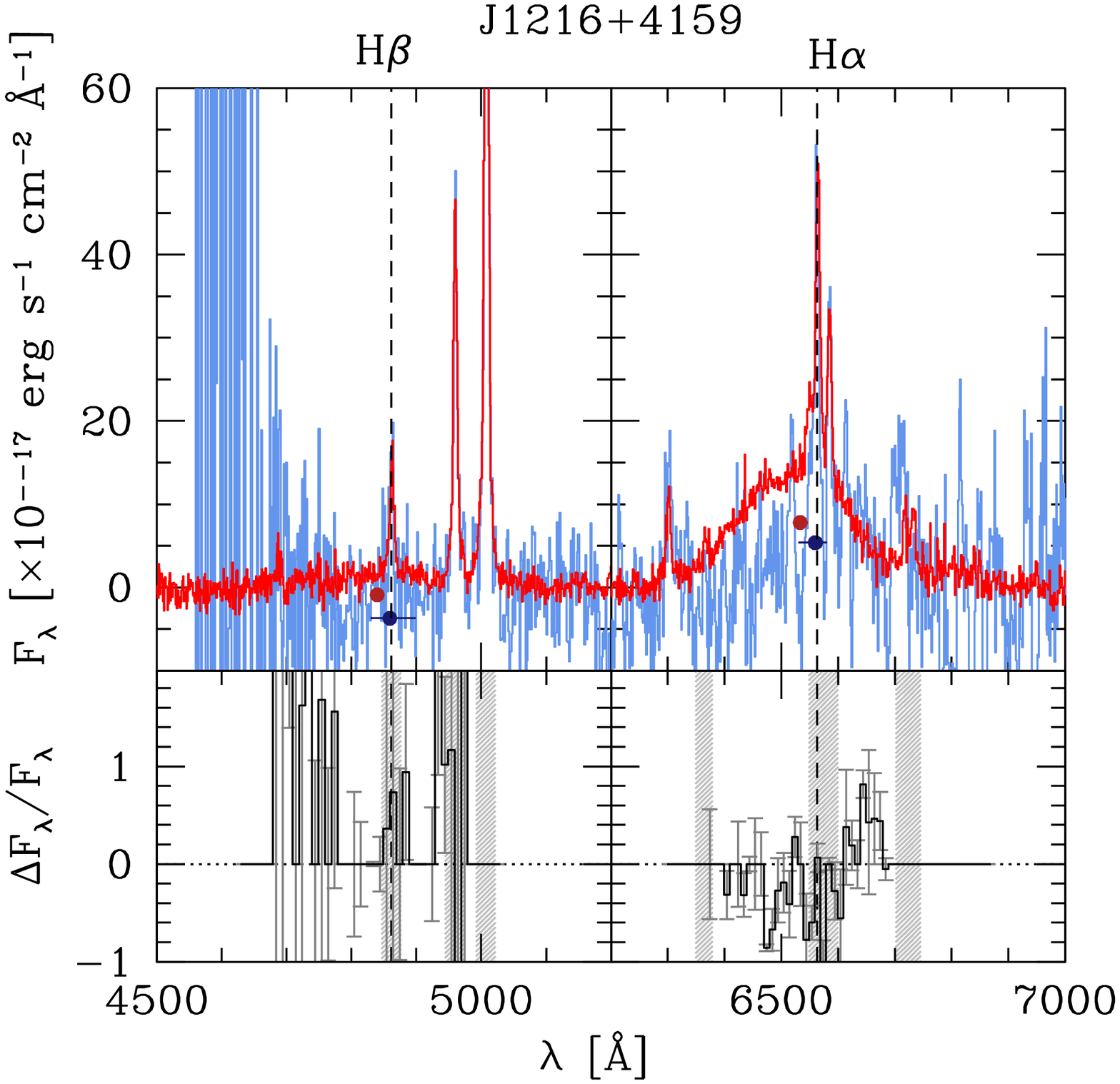} 
\includegraphics[width=0.3\textwidth]{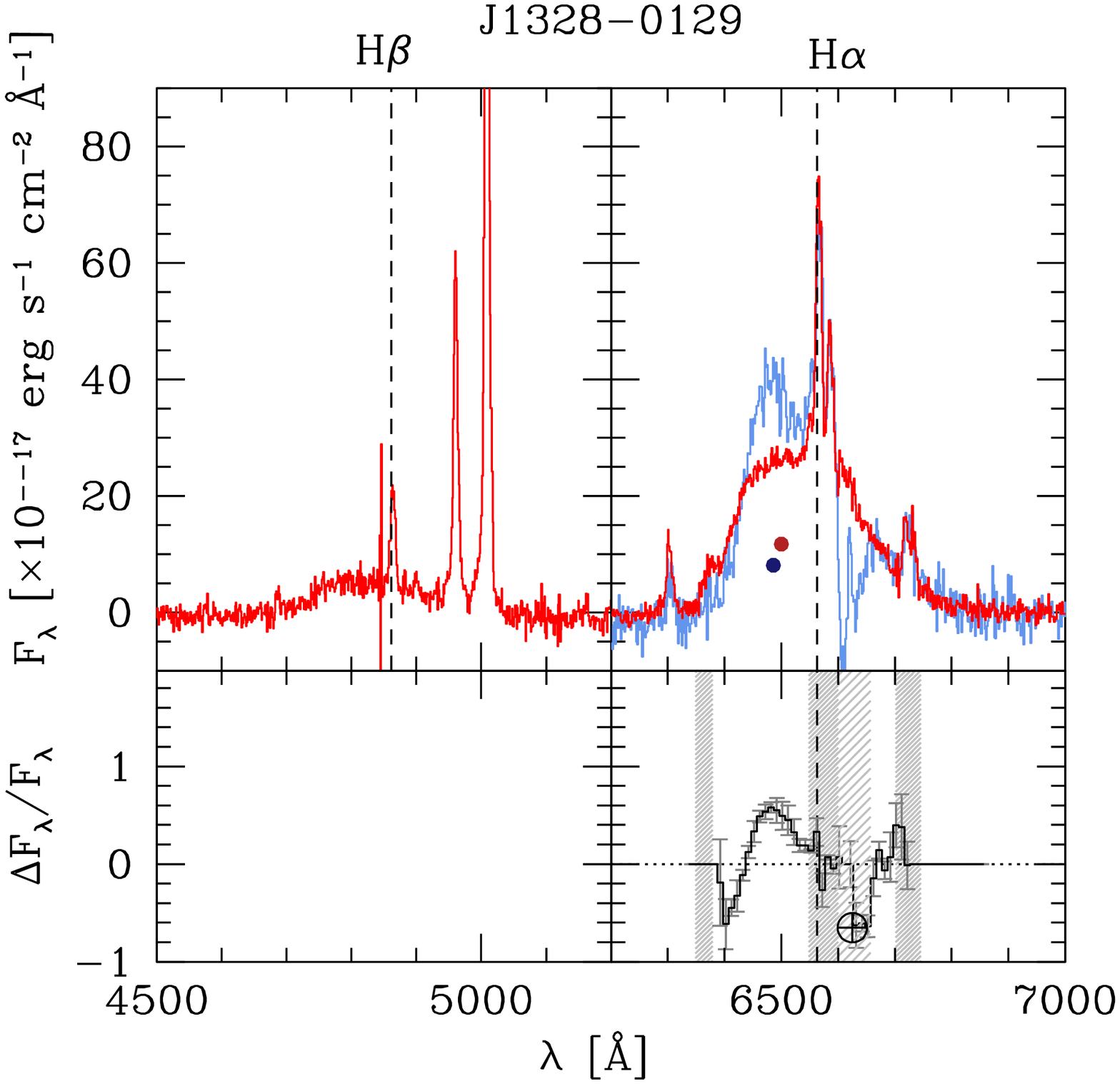} 
\includegraphics[width=0.3\textwidth]{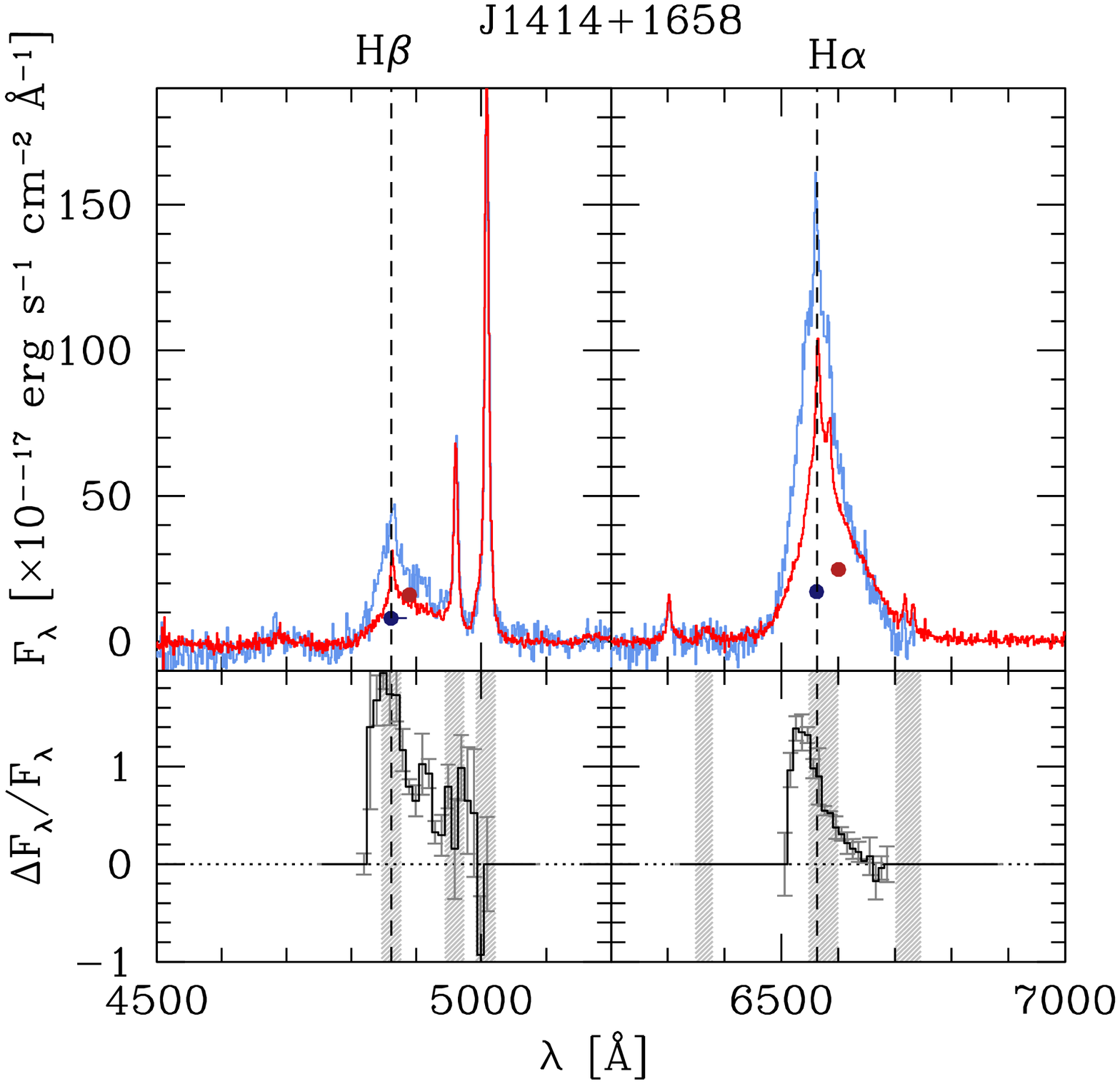} \\
\includegraphics[width=0.3\textwidth]{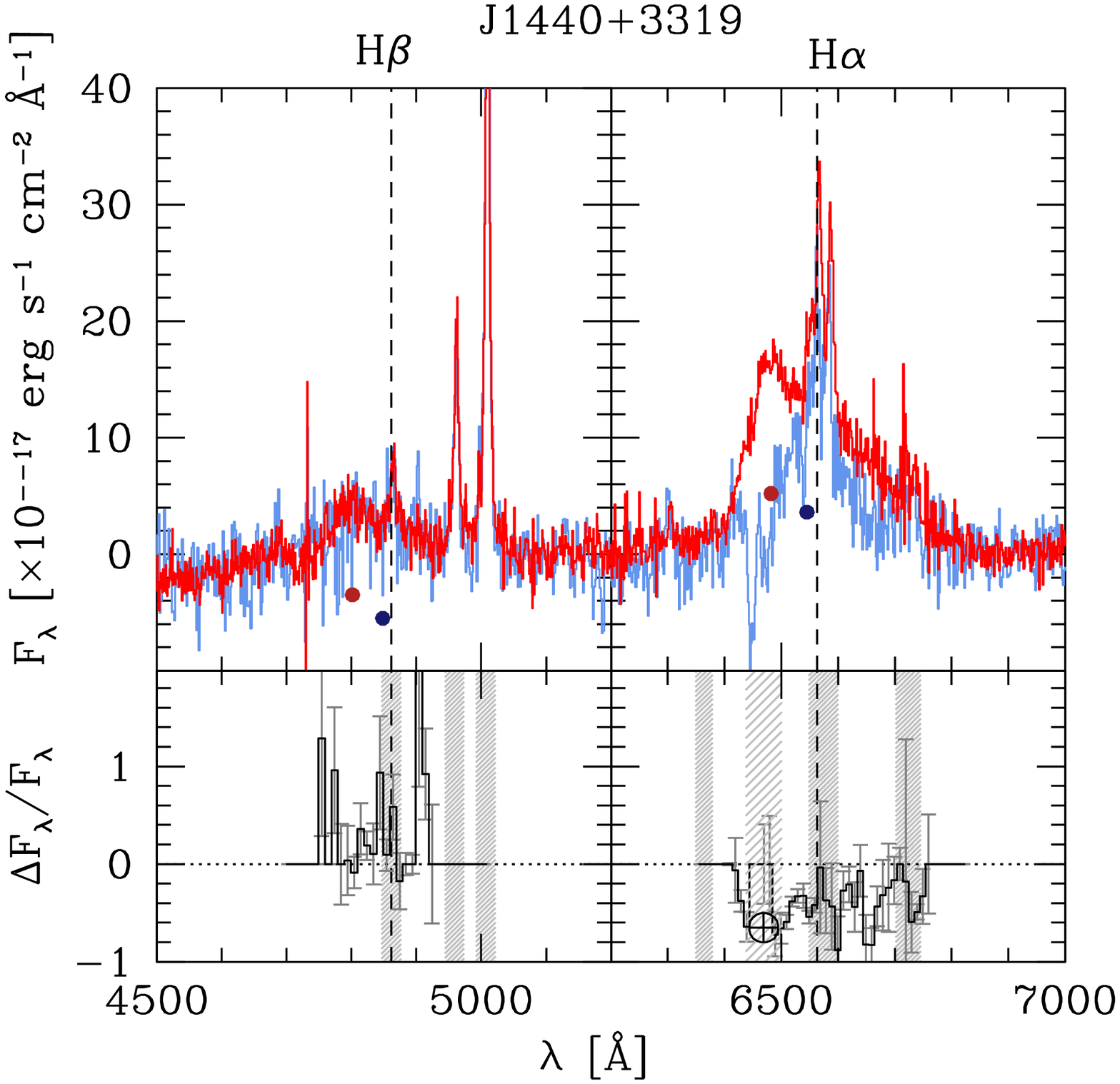} 
\includegraphics[width=0.3\textwidth]{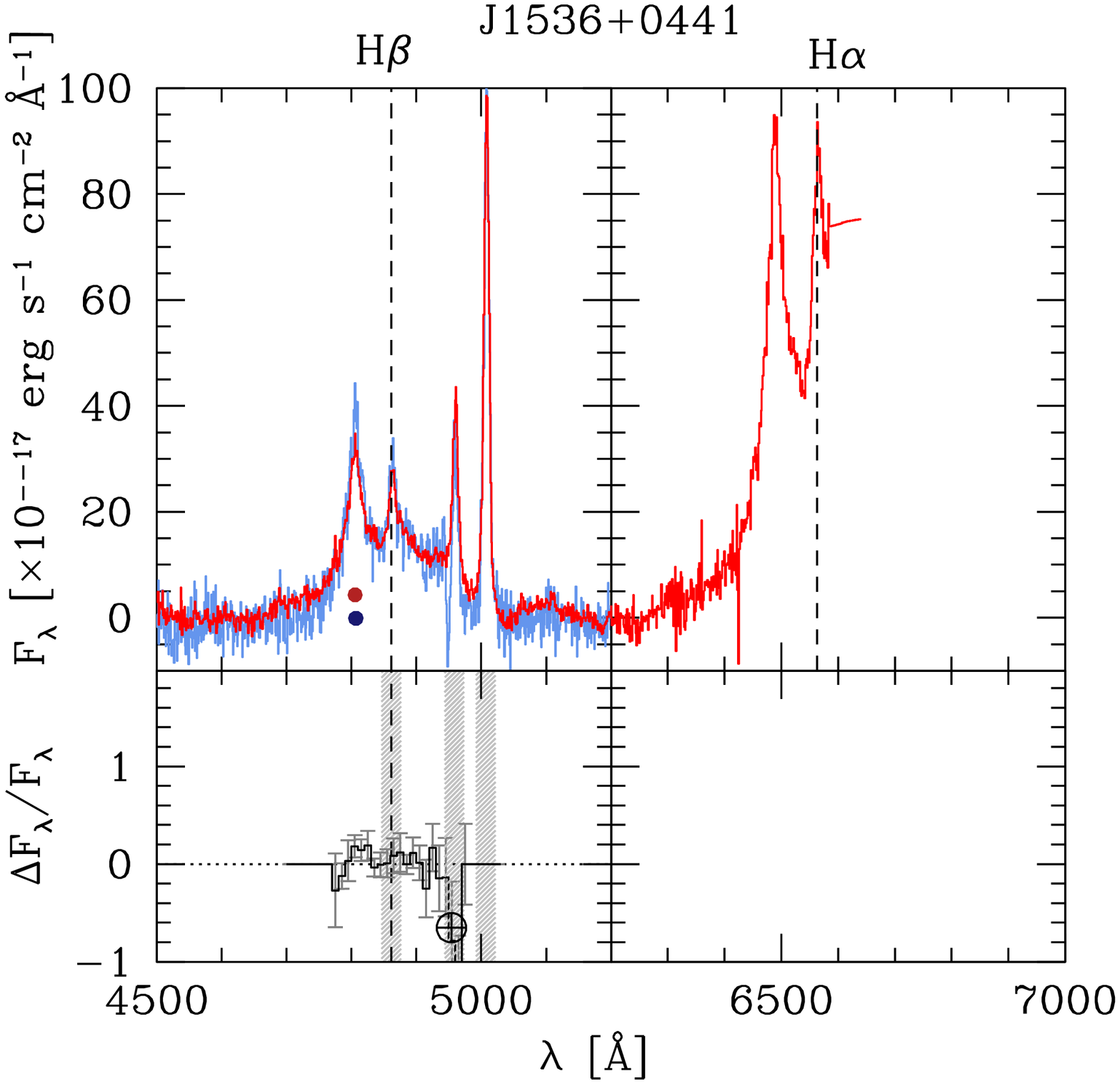} 
\includegraphics[width=0.3\textwidth]{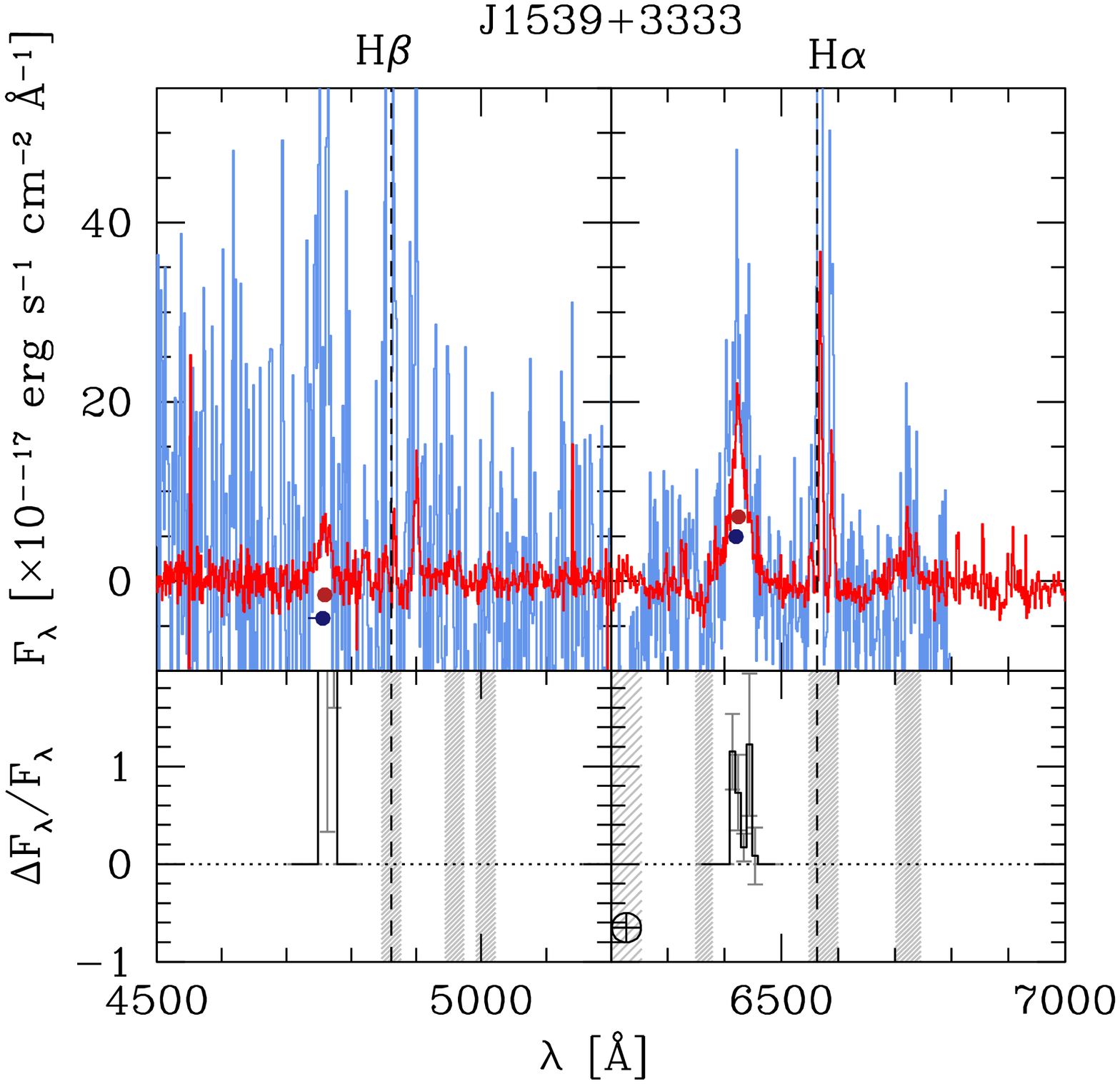} \\
\includegraphics[width=0.3\textwidth]{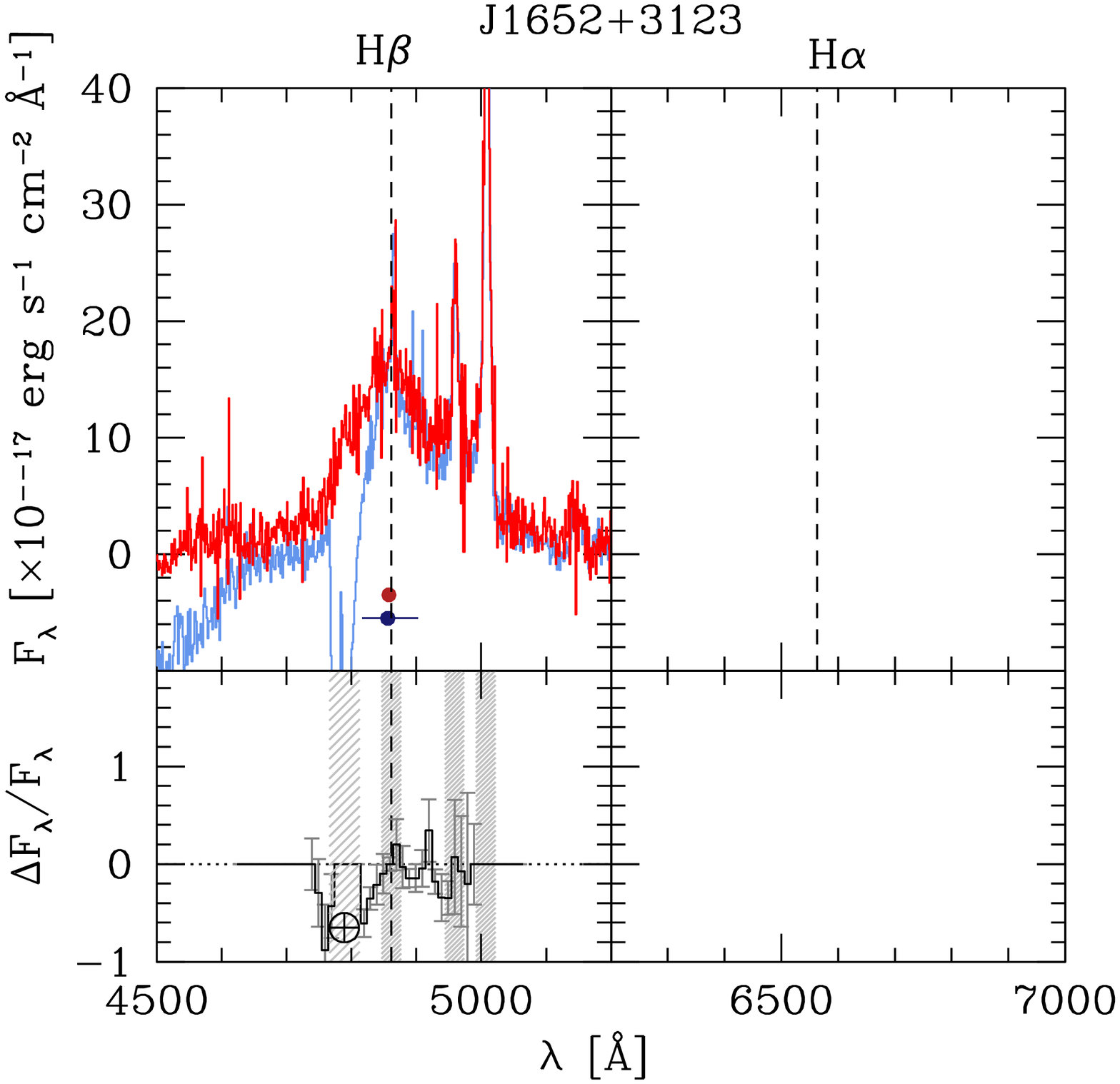} 
\includegraphics[width=0.3\textwidth]{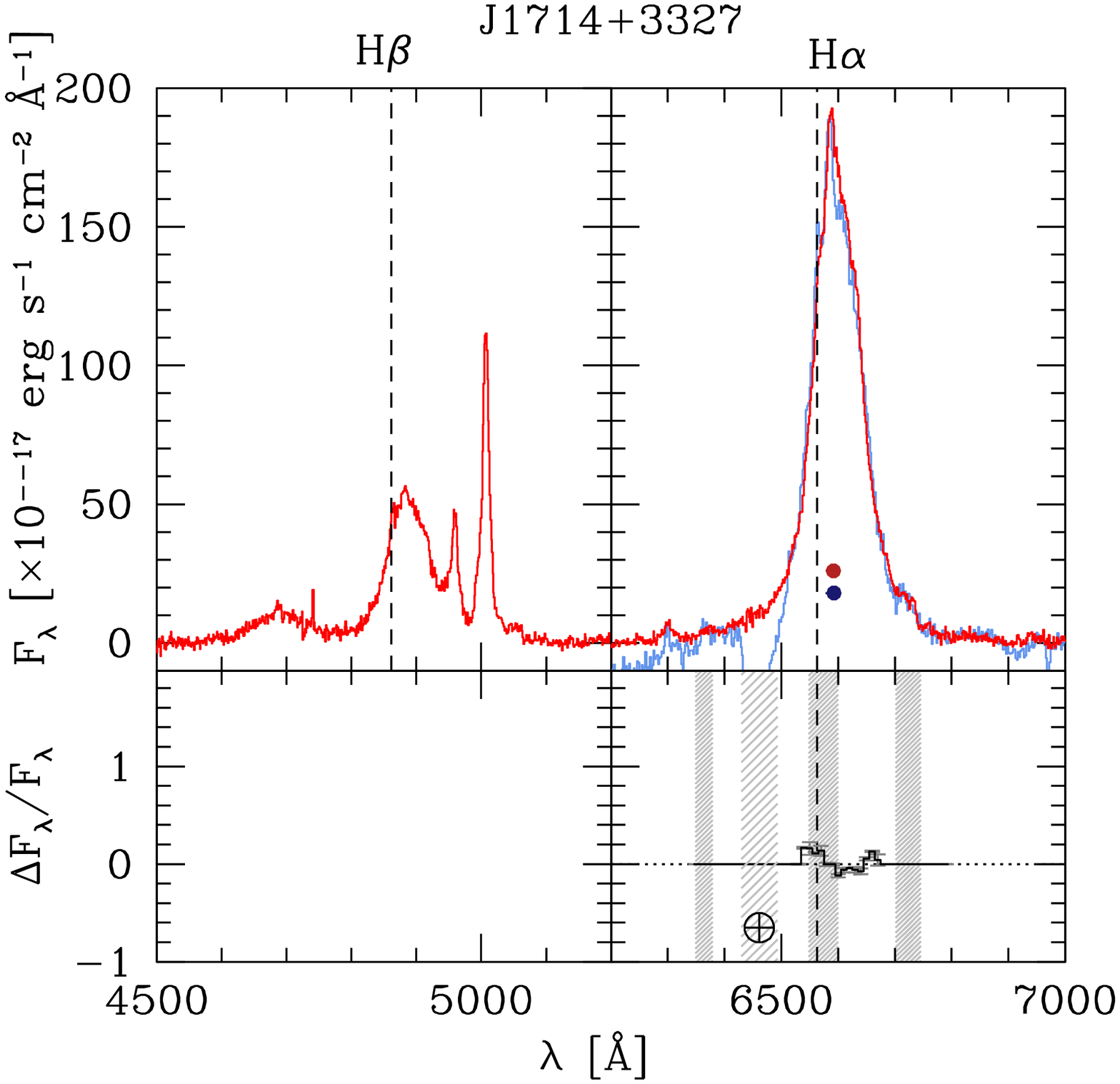} 
\caption{Continue from Figure \ref{fig_spc1}.}
\label{fig_spc3}
\end{figure*}

We note however that the changes in the peak wavelengths reported here may be 
part of a more general evolution in the line profile, which may not be 
associated with, e.g., a BHB. As a matter of fact, the general phenomenology 
of the line profile evolution is more complex.
For instance, the spectrum of J1414+1658 shows a clear change both in the peak 
wavelength and in the flux of the BLs. In J1328-0129, the evolution in the
BL shift is associated with the appearance of a new peak in the \Ha{} line 
profile at $\lambda \sim 6490$ \AA{}, while the base of the line profile
remained unchanged. The \Hb{} and \Ha{} lines in the J1207+0604 spectrum 
became more extended blue-wards, and the line profile more
boxy. In Figures \ref{fig_spc1}--\ref{fig_spc3}, we compare the 
continuum-subtracted line profiles of \Ha{} and \Hb{} observed in
SDSS data with those in our follow-up spectra, by plotting the normalized
difference between the observed BL profiles:
$[F_\lambda$(follow up) - $F_\lambda$(SDSS)]/$F_\lambda$(SDSS). In 
order to regularize this ratio, at the denominator we use the BL model of 
SDSS lines obtained in Section \ref{sec_spectra}, instead of the observed 
profile, and we bin into 10 \AA{}
intervals, to ease the visualization of small variations in regimes
of modest signal-to-noise. 

Significant line profile temporal variations are reported in 11 out of 32
sources. In 8 of them (J0012-1022, J0936+5331, J1117+6741, J1207+0604,
J1211+4647, J1328-0129, J1414+1658, J1440+3319), the profile of BLs changed 
in shape, with asymmetric features appearing or disappearing, shifts in the
peak wavelengths, etc. In the remaining three sources (J0221+0101, 
J0918+3156, J1216+4159) we observe a significant ($>$30\%) drop in the flux 
of BLs with no clear indication of a variation of the peak wavelength. Minor
($\sim10$\%) flux variations are also reported in additional 5 sources 
(J0932+0318, J0942+0900, J0946+0139, J1050+3456 and J1536+0441). 
A short description of the most prominent variations in the BL profiles
observed in this study is given in appendix \ref{sec_appendix}.

\begin{figure}
\includegraphics[width=0.99\columnwidth]{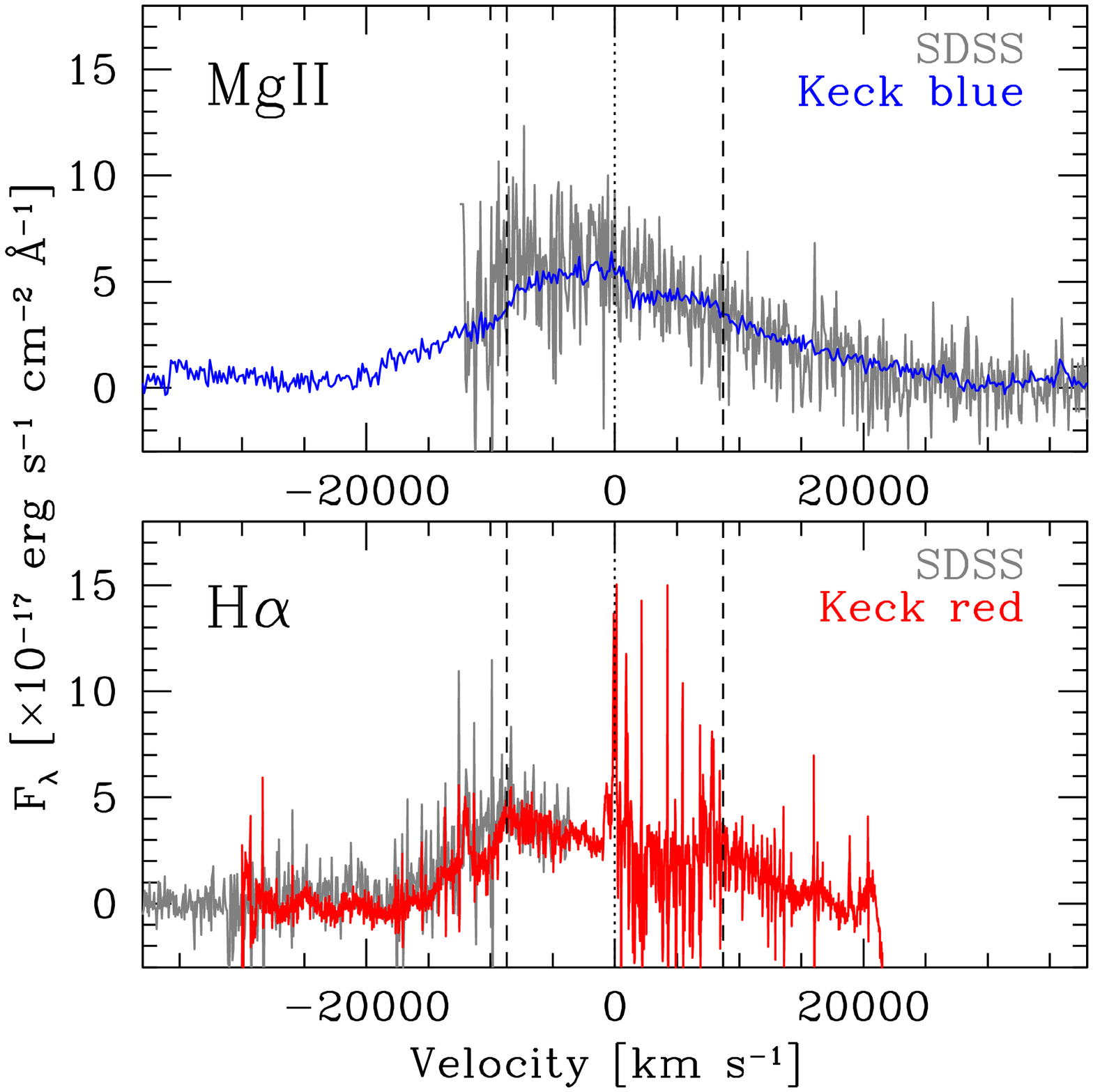}
\caption{The velocity intervals surrounding \Mgii{} and \Ha{} in
  J1000+2233. The broad \Hb{}
  and \Ha{} lines in the SDSS spectrum peak 8900 \kms{} bluewards of
  the narrow lines \citep[highlighted here with dashed, vertical 
  lines; see][]{decarli_4c2225}. Our follow-up Keck observations 
  expanded the observed range at both sides of the SDSS data,
  revealing an `M'-shaped profile of \Ha{}, and a bell-shaped
  \Mgii{} line profile, with significantly smaller peak shift. This 
  rules out the BHB hypothesis for this source.}
\label{fig_j1000}
\end{figure}

As final remarks: {\em (i)} For one object (J0155-0857) 
the poor quality of our follow-up observations hinders any conclusive
statements about the evolution of the line profiles. {\em (ii)}
J1000+2233 does not show any variation in the \Ha{} profile (the \Hb{} 
line is not covered by our follow-up). 
However, our Keck spectrum of this quasar covers for the first time the 
red wing of \Ha{} and the blue wing of \Mgii{} (see Figure \ref{fig_j1000}).
The former shows an `M'-shaped profile, while the latter is fairly bell-shaped
with significant smaller velocity shift compared with the blue peak of 
Balmer lines, suggesting that this sources is a DPE. 


\section{Discussion \& Conclusions}\label{sec_conclusions}

In this paper we investigate the nature of BHB candidates selected based 
on velocity shifts between broad and narrow lines in their SDSS spectra. 
According to the profiles of BLs, we divided the sample in four classes:
BHB candidates, showing fairly symmetric line profiles and big ($>$1000
\kms{}) velocity shifts; DPEs, with very broad lines, possibly showing
double-horned profiles; objects with Asymmetric lines; and other sources,
showing irregular line profiles, or modest shifts ($\sim 1000$ \kms{}) 
between BLs and NLs. 

The analysis of the SDSS discovery spectra reveals that:\\
 (1) BHB candidates cover a wide range in terms of broad and
 narrow line luminosities, widths, and virial estimates of the
 BH mass. DPEs tend to show very bright and broad lines (yielding
 large virial BH mass estimates); the only exception being the BHB/DPE
 candidate J1536+0441 \citep{boroson09}, which shows a modest
 FWHM for the broad component of \Hb{}. Asymmetric BLs tend show
 bright BL luminosities, and relatively high \Oiii{} 
 widths (FWHM$_{\rm [OIII]}\sim500$ \kms{}). Remaining sources tend to 
 have intermediate
 to low line luminosities and virial BH masses $<10^9$ 
 \Msun{}. Some of these sources show high \Ha{}/\Hb{} flux ratios in the
 broad components of their lines.
 In a few cases, NLs are significantly broadened (FWHM$_{\rm [OIII]}$ 
 $>$ 800 \kms{}).\\
 (2) For the majority of the sources, ionization properties in the 
 NL region are consistent with what typically observed in quasars. In 
 particular, hard radiation fields are needed to explain the generally
 high \Nev{}/\Neiii{} ratios observed in our sources. Only a handful 
 of targets (all classified as `BHB', `Asymmetric' or hybrid candidates)
 have modest or no \Nev{} emission, and may not require
 AGN-like ionization conditions. \\
 (3) Two sources (J0927+2943 and J1539+3333) show two sets of narrow
 lines, both at $z_{\rm NL}$ and at $z_{\rm BL}$. Another source
 (J1010+3725) shows broadened, multi-component narrow lines.
 All the remaining sources show no NLs at the redshift of the BLs.
 Ionization conditions of the secondary set of NLs in J1010+3725 and 
 J1539+3333 suggest modest or no AGN contribution; this is not the case
 for J0927+2943, which shows \Nev{} emission at both $z_{\rm NL}$
 and $z_{\rm BL}$.

By comparing the SDSS spectra with our second epoch spectra we found 
that:\\
 (1) 16 out of 32 objects did not undergo any significant evolution 
 in their broad line profiles or flux over a (rest-frame) timescale 
 of 2--10 yr. \\
 (2) Minor flux variations are reported in 5 targets, with no
 appreciable change in the line profile. \\
 (3) Three sources (J0221+0101, J0918+3156, J1216+4159) show conspicuous
 variations in the BL fluxes ($>30$\%), but with no clear change in the 
 shape of the line profiles.\\
 (4) Eight sources show significant evolution in the light profile, in 
 diverse forms. In particular, the BL peak wavelength shifted in 5 sources, 
 although such an evolution may be ascribed to a more general change in the 
 shape of the line profile.\\
 (5) Some degree of evolution in the line profiles is common in objects with 
 asymmetric lines (4 out of 5) and in `Other' sources (8 out of 15), while
 is less frequent in BHB candidates (including hybrid classifications, 
 only 3 sources out of 9 show some modest flux variation, and none of them
 show any significant shift in the BL peak wavelength) and in DPEs (1 out
 of 3). All the shifts in the BL peaks are associated with a general change 
 of the whole BL profile, rather than with a `rigid' shift of the line.

The absence of a clear shift in the majority of the objects (in particular, 
in those identified as BHB candidates) does not rule out a BHB explanation. 
Small to negligible differences in the peak position are expected for 
various plausible configurations, in particular if the binary orientation 
is such that maximizes the projected velocity shift (i.e., minimizes the 
acceleration component along the line of sight; see Figure \ref{fig_velevo},
{\em left}). On the other hand, fast evolution of the Balmer lines (likely 
associated with boxy profiles) is one of the predicted signatures of close 
and short period BHBs \citep{bogdanovic08, shen10}. It is still unclear 
whether the variability observed in our sample is associated with
BHBs or with bright spots in the BL region of `normal' AGN. 
Since changes in the BL peak wavelengths seem to be exclusively 
associated with dramatic fluctuations in line luminosities, with
relatively modest peak velocity changes, the BHB scenario is likely disfavoured
for these sources. Theoretically, in the BHB scenario the luminosity of the 
secondary is expected to evolve over a timescale comparable with the orbital 
period of the binary
\citep[][although fast fluctuations in the BH lightcurves on very
short timescales would not be temporarly resolved by the numerical
simulations cited above]{artymoviwicz96,hayasaki08,roedig11,sesana12,roedig12}.
New observations in 1--5 yr time lags will allow us to assess whether the
variations observed so far are transient features (as expected in the bright spot
scenario) or follow a regular evolution (as predicted in the BHB hypothesis).

\section*{Acknowledgments}

We thank Mike Eracleous, Bradley M. Peterson and Brent Groves for useful 
discussions on the AGN phenomenology.
We thank the Ulli Thiele, Santos Pedraz, and the astronomers in Calar Alto
for the precious support in the execution of the observations.
Support for RD was provided by the DFG priority program 1573 ``The physics
of the interstellar medium''. Support for MF was provided by NASA through Hubble 
Fellowship grant HF-51305.01-A awarded by the Space Telescope Science Institute, 
which is operated by the Association of Universities for Research in 
Astronomy, Inc., for NASA, under contract NAS 5-26555. 
Based on observations collected at the German-Spanish Astronomical Center,
Calar Alto, jointly operated by the Max-Planck-Institut f\"{u}r Astronomie
Heidelberg and the Instituto de Astrofisica de Andalucia (CSIC).
This research has made use of the NASA/IPAC Extragalactic Database (NED)
which is operated by the Jet Propulsion Laboratory, California
Institute of Technology, under contract with the National Aeronautics
and Space Administration. Funding for the Sloan Digital Sky Survey (SDSS)
has been provided by the Alfred P. Sloan Foundation, the Participating
Institutions, the National Aeronautics and Space Administration, the
National Science Foundation, the U.S. Department of Energy, the Japanese
Monbukagakusho, and the Max Planck Society. The SDSS Web site is
http://www.sdss.org/. The SDSS is managed by the Astrophysical Research
Consortium (ARC) for the Participating Institutions. The Participating
Institutions are The University of Chicago, Fermilab, the Institute for
Advanced Study, the Japan Participation Group, The Johns Hopkins University,
the Korean Scientist Group, Los Alamos National Laboratory, the
Max-Planck-Institute for Astronomy (MPIA), the Max-Planck-Institute for
Astrophysics (MPA), New Mexico State University, University of Pittsburgh,
University of Portsmouth, Princeton University, the United States Naval
Observatory, and the University of Washington.

\appendix
\section{Notes on individual targets}\label{sec_appendix}

Hereafter we briefly describe the changes observed in sources showing
prominent time variability in their line profiles:

  {\it J0012-1022} ---
  This object shows a dramatic change in the line profile. The knee observed
  in the red wing of the line in SDSS data disappeared in our follow-up
  observations (in a rest-frame time lag of $\approx 8$ yr), while the 
  asymmetric blue peak became brighter and shifted blue-wards. Such an 
  evolution is not confirmed in the \Hb{} follow-up observations reported by 
  \citet{eracleous12}, although the changes may be lost due to the 
  intrinsically fainter line, the confusion with blended \Feii{} and \Oiii{}
  emission, and the slightly lower signal-to-noise of their observations.

  {\it J0936+5331} ---
  The peak of the broad asymmetric line is significantly reduced (but still 
  visible as a knee in the red wing of the lines, roughly at the same 
  wavelength). The drop in the peak flux was already reported by 
  \citet{eracleous12}, their spectrum dating back to about 1 yr before ours.
  The brightest part of the line profile is now consistent with the core of 
  the line, closer to $z_{\rm NL}$. 

  {\it J1117+6741} ---
  The profile of broad \Ha{} evolved slightly, producing a shift in 
  the centroid of the line towards shorter wavelengths. This is
  not clearly observed for \Hb{} due to the significantly lower
  line flux.

  {\it J1207+0604} ---
  The BL peaks are 40\% fainter in our second epoch observations, and the
  blue wings of the line profile are slightly brighter. As a result, the 
  line profile appears boxier, and the line peak shifts blue-wards.

  {\it J1211+4647} ---
  The broad \Hb{} line became $\sim 10$\% fainter and shifted slightly 
  redwards, closer to the narrow \Hb{} line. This source appears also in the
  sample by \citet{eracleous12}, but no second epoch spectrum was collected 
  in their follow-up campaign for this source. 

  {\it J1328-0129} ---
  The broad \Ha{} line profile of J1328-0129, as observed in the SDSS data, 
  is boxy and blue-shifted of $\approx 3100$ \kms{} compared with the NLs. 
  Our second epoch spectrum reveals an increase of the flux emission
  of the blue wing, which now shows a clear peak at $\sim 7470$ \AA{}.
  Note that our conclusions are not affected by the atmospheric absorption 
  feature in the red wing of the line.

  {\it J1414+1658} ---
  The Balmer lines in this quasar became significantly brighter 
  (by a factor $\gsim 2$) and shifted blue-wards, closer to $z_{\rm NL}$.

  {\it J1440+3319} --- 
  The broad component of \Ha{} became fainter (by 30-50\%). In particular, the
  asymmetric peak observed in the SDSS spectrum seems disappeared (although our
  \Ha{} observations are affected by atmospheric absorption at those wavelengths,
  and the \Hb{} emission has modest signal-to-noise ratio).

\label{lastpage}

\end{document}